\documentclass{jfm-arxiv}

\usepackage{graphicx}
\usepackage{newtxtext}
\usepackage{newtxmath}
\usepackage{natbib}
\usepackage{hyperref}
\hypersetup{
    colorlinks = true,
    urlcolor   = blue,
    citecolor  = blue,
    linkcolor= blue,
}

\usepackage{bm}

\DeclareMathAlphabet{\mathsfit}{T1}{\sfdefault}{\mddefault}{\sldefault}
\SetMathAlphabet{\mathsfit}{bold}{T1}{\sfdefault}{\bfdefault}{\sldefault}

\newcommand{\ie}{i.e.}
\newcommand{\Fig}{Figure~\ref}
\newcommand{\fig}{figure~\ref}

\newcommand{\figs}{figures~\ref}
\newcommand{\Eq}{Eq.~\eqref}

\newcommand{\xx}[1]{\textcolor{black}{#1}}

\newcommand{\bn}{\bm{n}}

\title{Dissipation and microstructure in sheared active suspensions of squirmers}

\author{Zhouyang Ge\aff{1}
  \corresp{\email{zhoge@nju.edu.cn}},
 \and Gwynn J.~Elfring\aff{2}
  \corresp{\email{gelfring@mech.ubc.ca}}
  }

\affiliation{
\aff{1}School of Advanced Manufacturing Engineering, Nanjing University, Suzhou 215163, China
\aff{2}Department of Mechanical Engineering and Institute of Applied Mathematics, University of British Columbia, Vancouver V6T 1Z4, BC, Canada
}

\begin{document}

\maketitle

\setlength{\unitlength}{1cm}

\begin{abstract} 

We study the energy expenditure and structural correlations in semi-dilute to concentrated suspensions of squirmers using active fast Stokesian dynamics simulations. 
Specifically, we simulate apolar active suspensions of squirmers, or `shakers,' and show that shear enhances the total dissipation but reduces the relative viscosity for both puller- and pusher-type shakers.
At low shear rates where activity dominates,
\xx{hydrodynamic interactions increase the active dissipation, more so for pullers at higher volume fractions, whereas pushers exhibit a higher viscosity,} 
in contrast to bacterial suspensions displaying a `superfluid' transition.
At high shear rates where shear dominates, pullers and pushers behave effectively as passive spheres, generating negative normal stress differences due to shear-induced collision.
Remarkably, the rate-dependent rheological responses are accompanied by unusual microstructural signatures of an enhanced nematic order and anisotropic pair correlation, both of which contribute to a higher viscosity under shear.
Further simulations of self-propelled, neutral squirmers exhibit similar but weaker shear-thinning, highlighting the importance of activity over motility, underpinned by hydrodynamic interactions.
Overall, our results elucidate the interplay of internal activity and external flow on the dissipation and microstructure in sheared active suspensions of squirmers.

\end{abstract}

\section{Introduction}

Active suspensions differ from passive complex fluids in that the suspended particles can directly expend and dissipate energy.
This energy expenditure often results in particle \emph{motility}, as in the case of swimming microorganisms \citep{Ishikawa_2025perspective}; but it can also generate \emph{active stress}, leading to hydrodynamic instabilities and large-scale chaotic flows even without any external force \citep{Alert2022review}.
Given the prevalence of active particles in nature, the rich phenomena they display collectively, and their potential applications in environmental and biomedical technologies \citep{appl_environ2018, appl_medical2020}, the field of active suspensions, or \emph{active matter} more broadly, has excited much research from diverse disciplines \citep{koch_subramanian_review2011, marchetti_review2013, Saintillan2018, Vrugt2025metareview}.

One area that has received a lot of attention recently is the \emph{rheology} of active suspensions \citep{Saintillan2018}.
To appreciate its novelty, consider the rheology of passive suspensions first.
For a dilute suspension of rigid spheres under simple shear, \citet{Einstein1905viscosity, Einstein1911correction} showed that the relative suspension viscosity, $\eta_r$, increases linearly with the particle volume fraction, $\phi$, as $\eta_r = 1 + 2.5\phi$.
At higher concentrations ($\phi \gtrapprox 5\%$), the relationship becomes nonlinear due to particle interactions \citep{batchelor1970stress, guazzelli_pouliquen_2018}; however, one generally expects $\eta_r$ to increase with $\phi$ because the suspension has more solid content at higher $\phi$.
A striking observation in certain active suspensions, such as suspensions of {\it E.~coli} bacteria, is that $\eta_r$ can \emph{decrease} with $\phi$, and even \emph{vanishes} beyond a critical $\phi_c$, provided that the shear rate is small enough \citep{Saintillan2018}.
Indeed, various experiments have demonstrated the viscosity reduction and remarkable `superfluid' transition in bacterial suspensions  \citep{Sokolov2009, Gachelin2013, lopez2015turning}, probing the effects of concentration, shear rate, confinement, etc., as well as possible correlations of the mechanical and dynamical responses \citep{Guo_PNAS_2018, liu2019rheology, Martinez2020, Chui2021rheology}.
On the other hand, measurements of algal suspensions, which differ from bacteria in both swimming mechanism and shape, have shown the \emph{opposite} effect: $\eta_r$ increases with $\phi$ more than passive particles \citep{Rafai2010, Mussler2013}.
Clearly, there is a coupling between activity and flow, driving the fluid away from equilibrium and leading to complex rheology. 

Many theories have been proposed to predict or explain the rheological behaviors of active suspensions, starting from phenomenological descriptions for active liquid crystals \citep{Hatwalne2004, Cates2008active, Giomi2010active, Loisy2018}, to continuum kinetic theories for active rod-like particles \citep{Haines2009theory, saintillan2010dilute, saintillan2010extension, gyrya2011effective, Nambiar2017stress, Nambiar2019stress}, self-propelled spheres \citep{Takatori2017}, and autophoretic Janus spheres \citep{Traverso_Michelin_2022, Gaspard2026}.
Among them, the kinetic theory for rods appears to best describe the rheology of bacterial suspensions, as suggested by a recent experiment \citep{Martinez2020}.
The basic mechanism is that bacteria tend to have an orientation under shear such that their activity-induced flow \emph{assists} the external shear flow, thus reducing the effective viscosity.
In the dilute regime where particles do not interact, this effect scales linearly with $\phi$ (just like in the Einstein viscosity), thus the vanishing $\eta_r$ at a critical $\phi_c$.
However, the dilute assumption may become less accurate as $\phi \to \phi_c$, as bacteria tend to swim collectively in the superfluid-like state \citep{Guo_PNAS_2018, Martinez2020}.
Furthermore, the simple argument of flow-induced alignment does not apply to spherical active particles, unless some other mechanisms are invoked \citep{Rafai2010, Mussler2013}.
Therefore, it is necessary to go beyond the dilute regime and incorporate the effects of particle interactions to gain an improved and more quantitative understanding of the rheology of active suspensions.

Numerical simulations able to resolve the dynamics of individual particles are well-suited for the above purpose; yet, there are surprisingly few studies that have examined the rheology of active suspensions in the non-dilute regime.
Using a modified Stokesian dynamics method, Ishikawa and coworkers \citep{ishikawa2007rheology, ishikawa2021rheology, Ishikawa_Brumley_Pedley_2025} simulated semi-dilute and dense suspensions of spherical `squirmers' in simple shear and Poiseuille flows. 
Their studies tended to focus on so-called bottom-heavy squirmers, which were subject to a gravitational torque and could display unconventional rheological behaviors depending on the directions of the flow and gravity.
For non-bottom-heavy squirmers, they showed that activity could increase the relative viscosity and lead to shear-thinning in the dense regime \citep{ishikawa2021rheology}; though the simulations assumed that the squirmers lie perfectly on a monolayer, which is dynamically unstable.
Using a lattice Boltzmann method, \citet{Pagonabarraga2013} simulated squirmer suspensions in a narrow slit between two walls.
The authors found that the relative viscosity could vanish for both `puller'-type (\eg algae) and `pusher'-type (\eg bacteria) particles at small shear rates, which they attributed to the development of shear bands near walls. 
We note that such viscosity reduction of pullers has not been observed experimentally.
Furthermore, the setup of a three-dimensional narrow slit and a two-dimensional orientation in the shear plane is problematic: unless there is an exact symmetry in the vorticity direction, which is unstable, hydrodynamic interactions would rotate the squirmers off the shear plane. 

The scarcity of rigorous hydrodynamic studies on the rheology of, and more generally, \emph{dissipation} in concentrated active suspensions motivates us to investigate this problem using active fast Stokesian dynamics simulations \citep{fiore2019fast, elfring2022active}.
Specifically, we focus on \emph{apolar} active suspensions of squirmers, which we have previously studied in the absence of any external flow \citep{Ge_Elfring_2025} and under simple shear at various shear rates and volume fractions ($\phi \in [0.1, 0.4]$ mostly) \citep{Ge_Brady_Elfring2025}.
The choice of apolar active suspensions is to both isolate the effect of activity and reduce the number of parameters, as will become clear later. 
Systematic analyses show that shear enhances the total dissipation but reduces the relative viscosity, corresponding to a shear-thinning rheology, for both puller- and pusher-type squirmers.
The rate-dependent rheological responses are also accompanied by unusual microstructural signatures in terms of the global nematic order and local pair correlation, which we characterize in detail.
Further simulations of suspensions of self-propelled, neutral squirmers at $\phi=10\%$ exhibit similar but weaker shear-thinning, in contrast to the prediction of a kinetic theory neglecting particle interactions \citep{Takatori2017}.
Overall, our results illustrate the coupled effects of activity and shear on the dissipation and microstructure in squirmer suspensions, highlighting the importance of activity over motility, underpinned by hydrodynamic interactions.

The rest of the paper is organized as follows.
In \S \ref{sec:method}, we briefly describe the mathematical formulation and numerical setup of our simulations, \xx{including the energy balance and the dissipation in suspensions of shakers (\S \ref{sec:pre} and Appendix \ref{sec:edi_int}).
In \S \ref{sec:results}, we present the simulation results, starting from the dissipation (\S \ref{sec:edi}), then the rheology (\S \ref{sec:rheo}), and finally the microstructure (\S \ref{sec:struct}) to interpret the observations.}
We conclude and offer our perspectives in \S \ref{sec:conclusion}.

\section{Models and methods}
\label{sec:method}

\subsection{Active Stokesian dynamics}

We model active particles as squirmers, spheres with prescribed axisymmetric slip velocities \citep{lighthill1952, blake1971, Pedley2016}.
Specifically, we consider two types of squirmers with contrasting behaviors: (i) apolar squirmers, or `shakers,' that are individually immotile but collectively display hydrodynamic diffusion \citep{Ge_Elfring_2025}, and (ii) neutral squirmers, or active Brownian particles (ABPs) \citep{Romanczuk2012ABPs}, which self-propel and rotate spontaneously but barely interact hydrodynamically.
We simulate shakers (mainly) or ABPs in shear flow using an active version of the fast Stokesian dynamics (FSD) method \citep{fiore2019fast, elfring2022active}.
The mathematical formulation, algorithmic details, and numerical verification of the original and our modified FSD methods have already been extensively documented \citep{fiore2019fast, elfring2022active, Ge2022, Ge_Elfring_2025, Ge_Brady_Elfring2025}.
Since this work follows directly from our previous ones, only a brief summary of the governing equations and numerical setup is presented as follows.

Assuming inertialess particles suspended in a viscous fluid, the external force moments on any particle must be balanced by their hydrodynamic counterparts, which at low Reynolds numbers are linearly related to the velocity moments through the hydrodynamic resistance tensors.
For athermal squirmers under shear, this leads to \citep[note the different sign convention from][]{elfring2022active}
\begin{equation} \label{eq:active-sd}
    \mathbf U =  \mathbf U_\infty + \mathbf U_\text{s} + \mathsfbi{R}_\mathrm{FU}^{-1} \bm\cdot
    \big(\mathsfbi R_\text{FE} \bm: (\mathsfbi E_\infty + \mathsfbi E_\text{s}) + \mathbf F_\text{ext} \big),
\end{equation}
where the velocities ($\mathbf U, \mathbf U_\infty$, $\mathbf U_\text{s}$), strain rates ($\mathsfbi E_\infty, \mathsfbi E_\text{s}$) and force ($\mathbf F_\text{ext}$) are all $N$-tuples, with elements defined at the particle centers, e.g., ${\mathbf U} \equiv (\bm U_1 \ \bm U_2 \ ... \ \bm U_N)^T$, and ${\mathsfbi R}_\text{FU}$ and ${\mathsfbi R}_\text{FE}$ are the grand resistance tensors coupling the force and velocity moments of all $N$ particles.
In this compact notation, the velocities and forces defined on each particle include both the linear and angular components as in the original Stokesian dynamics formulation \citep{sd1988}.

Eq.~\eqref{eq:active-sd} shows that particle motion can be driven by external flow ($\mathbf U_\infty$ and $\mathsfbi E_\infty$), self-propulsion ($\mathbf U_\text{s}$), activity ($\mathsfbi E_\text{s}$), interparticle force ($\mathbf F_\text{ext}$), or a combination thereof.
Externally, we apply a simple shear with shear rate $\dot\gamma$ along the $x$ direction, resulting in an ambient velocity $\bm U_\infty(x,y,z) = (\dot\gamma y, 0, 0)$ that varies in the $y$ direction.
Internally, each squirmer may self-propel or rotate and induce a local strain in the fluid \citep{elfring2022active},
\begin{align} \label{eq:Es}
\bm U_\text{s} = \frac{2}{3}B_1 \bm p, \quad 
\bm \Omega_\text{s} = \frac{C_1}{a^3} \bm p, \quad 
\mathsfbi E_\text{s} = \frac{3B_2}{5a} \bigg(\bm p \bm p-\frac{1}{3} \mathsfbi I \bigg),
\end{align}
where $B_1$ and $B_2$ are the leading-order squirming coefficients,
$C_1$ is a coefficient associated with azimuthal slip,
$a$ is the particle radius,
$\bm p$ is the particle director,
and $\mathsfbi I$ is an identity tensor.
In principle, the slip velocity can also consist of higher-order velocity moments, but they are omitted here as done typically \citep{elfring2022active}.
Setting $B_1=0$ yields an individually immotile squirmer known as a \emph{shaker}.\footnote{We also set $C_1=0$ for shakers to emphasize the effect of hydrodynamic interactions on particle reorientation, as was done previously \citep{Ge_Brady_Elfring2025}.}
Although an isolated shaker cannot swim, a suspension of them stirs the fluid and exhibits collective motion \citep{Ge_Elfring_2025}.
For squirmers not subject to any external force or flow, the leading order hydrodynamic interactions arise from $\mathsfbi E_\text{s}$ if $B_2 \ne 0$: $B_2>0$ corresponds to \emph{pullers}, while $B_2<0$ \emph{pushers}; both are considered in this work. 
In addition to shakers, we also simulate neutral squirmers with an imposed self-propulsion speed, $U_\text{s}$, and rotational diffusion coefficient, $d_r$, satisfying
\begin{align} \label{eq:dr}
    \langle \bm \Omega_\text{s}(t) \bm \Omega_\text{s}(t^\prime) \rangle = 2d_r \delta(t-t^\prime) \mathsfbi I ,
\end{align}
where $\langle \cdot \rangle$ is an ensemble average, $\delta(t)$ is the Dirac delta, and $\bm \Omega_\text{s}$ is about an arbitrary axis (not necessarily $\bm p$) \citep{Fily2012}.
Such diffusion may be due to any internal mechanism of athermal origin, e.g., the tumble dynamics of swimming organisms \citep{Berg_book}.
Setting $B_2=0$ removes the hydrodynamic interactions between these squirmers at the leading-order we consider (if $\mathsfbi E_\infty= \bm 0$), making them equivalent to ABPs, a widely studied model for self-propelled particles, such as self-phoretic colloids \citep{Golestanian2009, Romanczuk2012ABPs}.  
Furthermore, to avoid near contact of the particles and thus introducing numerical difficulties, we impose a short-range repulsive force, $\bm F_\text{ext} = \bm F_0 \exp (-\kappa h)$, where $F_0$ is the maximal repulsion, $\kappa$ is an inverse characteristic length, and $h$ is the surface gap between the particles.
Physically, $\bm F_\text{ext}$ resembles the electrostatic repulsion between colloids suspended in liquids \citep{Israelachvili_book, mewis_wagner_book}.

\subsection{Stress calculations}
\label{sec:pre}

In a particle suspension, the bulk stress tensor, $\mathsfb\Sigma$, can be decomposed into fluid and particle stresses, $\mathsfb\Sigma_f$ and $\mathsfb\Sigma_p$, respectively \citep{batchelor1970stress}
\begin{align}\label{eq:Sigma}
   \mathsfb\Sigma  =  \mathsfb\Sigma_f + \mathsfb\Sigma_p  =
  -  \Pi_f  \mathsfbi I + 2\eta  \mathsfbi E_\infty  +  \mathsfb\Sigma_p  ,
\end{align} 
where $\Pi_f$ is the bulk fluid pressure, and $2\eta \mathsfbi E_\infty$ is the deviatoric Newtonian stress at the bulk rate of strain.
The last term, under the condition that the fluid and particle inertia are negligible, may be calculated as \citep{sd1988, elfring2022active}
\begin{align}\label{eq:Sigma_p}
   \mathsfb\Sigma_p  = -n \langle \mathsfbi R_\text{SU} \bm\cdot (\bm U - \bm U_\text{s} - \bm U_\infty) 
    - \mathsfbi R_\text{SE} \bm: (\mathsfbi E_\text{s} + \mathsfbi E_\infty)  + \bm x \bm F_\text{ext} \rangle ,
\end{align} 
where $n$ is the number density of the particles, $\mathsfbi R_\text{SU}$ and $\mathsfbi R_\text{SE}$ are the per-particle resistance tensors relating the particle stress to the local relative velocity ($\bm U - \bm U_\text{s} - \bm U_\infty$) and rate-of-strain ($-\mathsfbi E_\text{s} - \mathsfbi E_\infty$), respectively, $\bm x$ is the particle position, and $\langle \cdot \rangle$ averages over all particles.
Substituting \Eq{eq:active-sd} into \eqref{eq:Sigma_p}, we obtain
\begin{align}\label{eq:Sigma_p1}
   \mathsfb\Sigma_p  & = 
  n \big\langle \big(\mathsfbi R_\text{SE} - \mathsfbi R_\text{SU} \bm\cdot \mathsfbi R_\text{FU}^{-1} \bm\cdot \mathsfbi R_\text{FE} \big) \bm: (\mathsfbi E_\text{s} + \mathsfbi E_\infty)  
    - \mathsfbi R_\text{SU} \bm\cdot \mathsfbi R_\text{FU}^{-1}\bm\cdot \bm F_\text{ext}  - \bm x \bm F_\text{ext}  \big\rangle,
\end{align}
\xx{where we can identify a `hydrodynamic stresslet' due to the fluid,
\begin{align}\label{eq:S_H}
  \mathsfbi S_\text{H} \equiv (\mathsfbi R_\text{SE} - \mathsfbi R_\text{SU} \bm\cdot \mathsfbi R_\text{FU}^{-1} \bm\cdot \mathsfbi R_\text{FE} ) \bm: (\mathsfbi E_\text{s} + \mathsfbi E_\infty) - \mathsfbi R_\text{SU} \bm\cdot \mathsfbi R_\text{FU}^{-1}\bm\cdot \bm F_\text{ext}, 
\end{align}
and an `interparticle stresslet' due to non-hydrodynamic effects, $\mathsfbi S_\text{P} \equiv - \bm x \bm F_\text{ext}$.
Note that, $\bm F_\text{ext}$ also appears in $\mathsfbi S_\text{H}$.
}

\xx{From the bulk stress, the rate of \emph{external} dissipation can be calculated as \citep{batchelor1970stress}
\begin{align}\label{eq:edi_ext}
   \Phi_\text{ext}  \equiv  \mathsfb\Sigma \bm:  \mathsfbi E_\infty  
   = 2\eta ( \mathsfbi E_\infty \bm: \mathsfbi E_\infty )+ 
   n \langle \mathsfbi S_\text{H} + \mathsfbi S_\text{P} \rangle \bm: \mathsfbi E_\infty .
\end{align}
Here, $\Phi_\text{ext}$ measures the power supplied by external boundaries to deform a unit volume of the suspension at strain rate $\mathsfbi E_\infty$.
In simple shear, which is the case we consider, the above simplifies to 
\begin{align}\label{eq:edi_ext1}
   \Phi_\text{ext}  = \eta \dot\gamma^2 + n\langle S_{H,xy} + S_{P,xy} \rangle \dot\gamma ,
\end{align}
where we recognize $\Phi_f \equiv \eta \dot\gamma^2$ as the dissipation of a pure fluid at the same shear rate.
The ratio between $\Phi_\text{ext}$ and $\Phi_f$ thus describes the relative viscosity of the suspension,  
\begin{align}\label{eq:eta_r}
  \eta_r \equiv \frac{\Phi_\text{ext}}{\Phi_{f}} 
  = 1 + \frac{n}{\eta \dot\gamma}\langle S_{H,xy} + S_{P,xy} \rangle.
\end{align}
Clearly, $\eta_r$ depends on both shear and activity as $\mathsfbi E_\infty$ and $\mathsfbi E_\text{s}$ appear together in \Eq{eq:S_H}.
}

\xx{An important distinction between active and passive particles is that the former can directly dissipate energy due to the active strain, $\mathsfbi E_\text{s}$.
 In the framework of active Stokesian dynamics, this \emph{internal} dissipation can be expressed as
 \begin{align}\label{eq:edi_int}
   \Phi_\text{int}  \equiv n \langle \mathsfbi S_\text{tr} \bm: \mathsfbi E_\text{s} \rangle 
   =  n \langle \mathsfbi S_\text{H} \bm: \mathsfbi E_\text{s} \rangle - 2\eta \phi \langle \mathsfbi E_\text{s} \bm: \mathsfbi E_\text{s} \rangle,
\end{align}
where $\mathsfbi S_\text{tr}$ is the symmetric and traceless first moment of the traction on an active particle, which can differ from $\mathsfbi S_\text{H}$ if the surface velocity does not satisfy rigid-body motion \citep{elfring2022active}. 
For the shakers we consider, this results in a correction represented by the last term on the right-hand side of \Eq{eq:edi_int}; see Appendix \ref{sec:edi_int} for its derivation.
} 

\xx{Eq.~\eqref{eq:edi_int} shows that $\Phi_\text{int}$ can depend on shear.}
To set a reference for later comparisons, we denote the internal dissipation without shear as $\Phi_0 \equiv \Phi |_{\dot\gamma=0} = \Phi_\text{int} |_{\dot\gamma=0}$.
On dimensional grounds, $\Phi_0$ depends on the fluid viscosity $\eta$ and the activity time scale $\tau_a$ \xx{(fixing $\bm F_\text{ext}$)}.
A peculiar feature of ABPs is that \xx{an isolated ABP} does \emph{not} have any internal dissipation, which is of course an idealization ($\mathsfbi E_\text{s} \equiv \bm 0$).
In general, squirmers generate local strains as they swim, thus dissipating energy to the surrounding fluid.

\xx{The above shows that the total dissipation, $\Phi$, in an active suspension at steady state is balanced by two power inputs, one at the external driving boundaries, the other at the internal active surfaces.
Therefore, similar to $\eta_r$, we can define a relative dissipation, }
\begin{align}\label{eq:edi_r}
  \varepsilon_r \equiv \frac{\Phi}{\Phi_{f}} 
  = \eta_r + \xx{\frac{n}{\eta \dot\gamma^2} \langle \mathsfbi S_\text{H} \bm: \mathsfbi E_\text{s} \rangle - \frac{2\phi}{\dot\gamma^2} \langle \mathsfbi E_\text{s} \bm: \mathsfbi E_\text{s} \rangle},
\end{align}
characterizing the rate of total energy expenditure in the suspension.

Finally, before we proceed to the simulation setup and general results, we may consider a dilute suspension of shakers ($\phi \to 0$) to gain some intuition.
In the absence of any external force or torque, the stress an isolated shaker imparts to the fluid is given by \citep{elfring2022active}
\begin{align}\label{eq:stresslet}
  \mathsfb\Sigma_p = \xx{n \langle \mathsfbi S_\text{H} \rangle}
  = n \langle \mathsfbi R_\text{SE} \bm : (\mathsfbi E_\text{s} +\mathsfbi E_\infty) \rangle 
  =\frac{20}{3}n \pi \eta a^3 \langle \mathsfbi E_\text{s} +\mathsfbi E_\infty \rangle
  = 5\eta \phi \langle \mathsfbi E_\text{s} +\mathsfbi E_\infty \rangle.
\end{align} 
Without any ambient flow ($\mathsfbi E_\infty = \bm 0$), the total dissipation equals the internal one and is independent of the particle orientation,
\begin{align}\label{eq:edi-iso}
  \Phi_\text{iso} = \Phi_\text{int} |_{\dot\gamma=0} 
  = \xx{ n\langle \mathsfbi S_\text{H} \bm: \mathsfbi E_\text{s} \rangle - 2\eta \phi \langle \mathsfbi E_\text{s} \bm: \mathsfbi E_\text{s} \rangle}
  = \xx{3} \eta \phi  \langle \mathsfbi E_\text{s} \bm: \mathsfbi E_\text{s} \rangle
  = \xx{\frac{18}{25}} \frac{\eta\phi}{\tau_a^2},
\end{align} 
where $\tau_a \equiv a/|B_2|$ as a reminder.
With shear, the relative viscosity is 
\begin{align}\label{eq:visc_sq}
  \eta_r = 1 + \frac{5\phi}{\dot\gamma} \langle E_{\infty,xy} + E_{\text{s},xy} \rangle= 
  1 + \bigg[ 2.5+ \frac{3B_2}{\dot\gamma a} \langle p_x p_y \rangle \bigg] \phi.
\end{align} 
\Eq{eq:edi-iso} shows that an isolated pulling shaker dissipates the \emph{same} amount of energy as an isolated pushing shaker per unit time and volume.
Adding an external shear will not alter this equivalence, \emph{on average}, as spherical particles rotate under shear but do not have any preferred orientation ($\langle p_x p_y \rangle =0$), reducing \Eq{eq:visc_sq} to the Einstein viscosity for passive spheres. 
This is in contrast to elongated particles, such as {\it E.~coli} bacteria, which tend to align with the flow, thus breaking the puller-pusher symmetry and rendering non-Newtonian rheology \citep{Hatwalne2004, saintillan2010dilute}.

\subsection{Simulation setup}

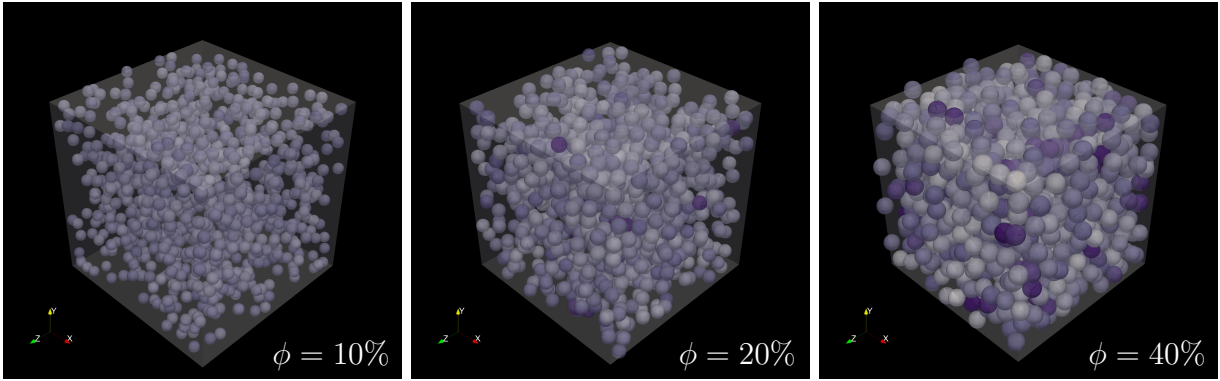
\begin{figure*}
  \centering
  \includegraphics[height=4.cm]{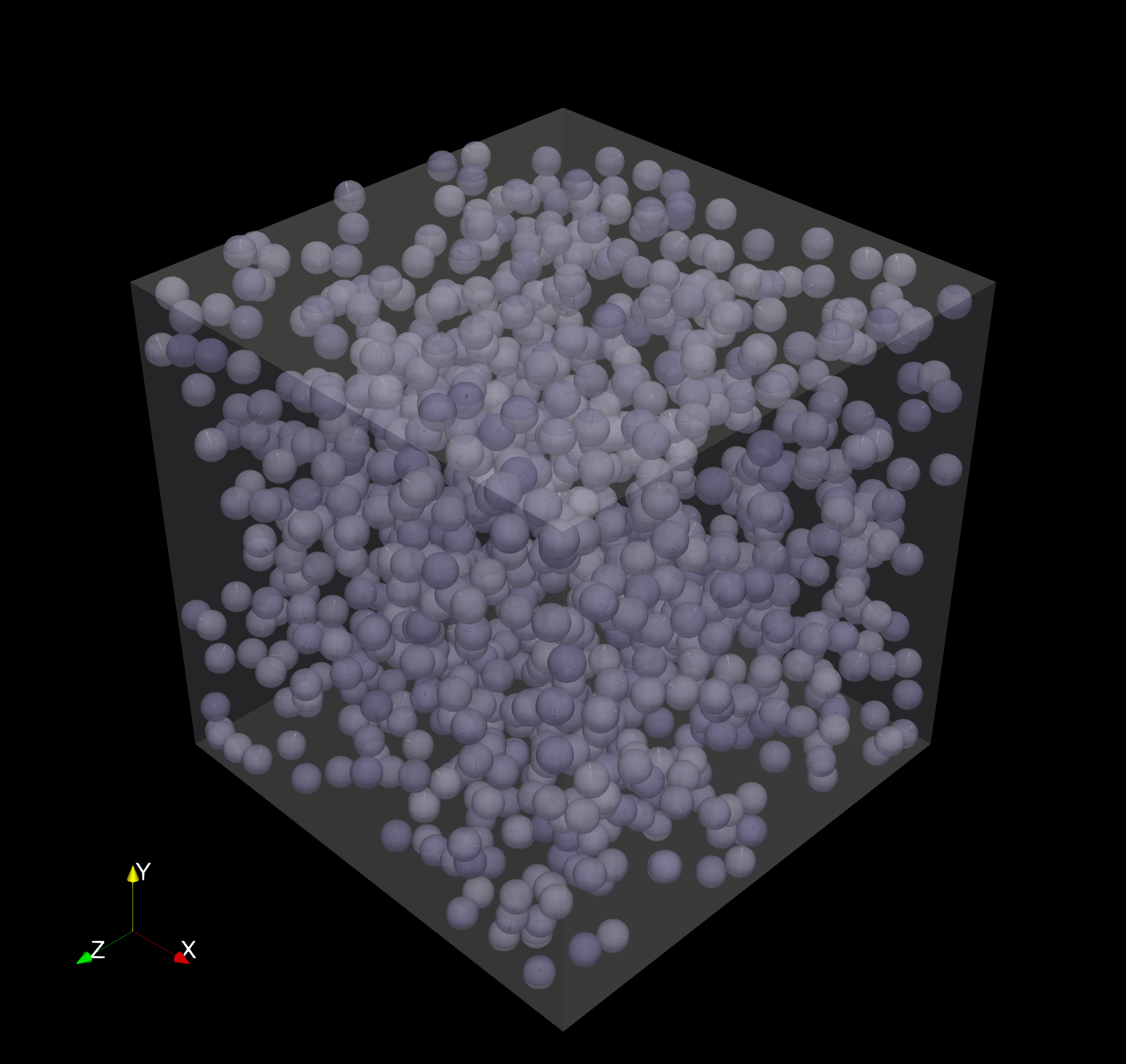} 
  \includegraphics[height=4.cm]{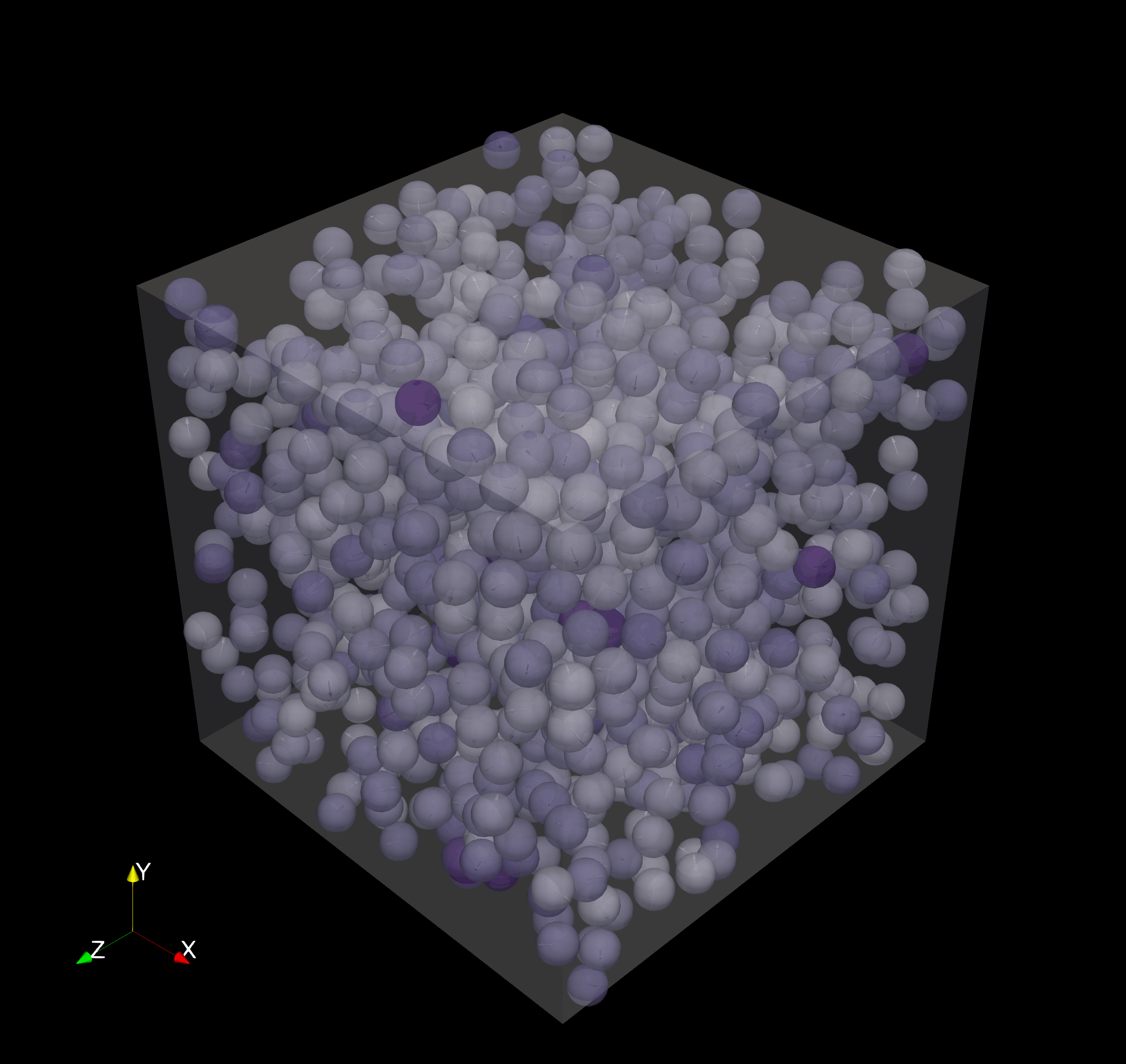} 
  \includegraphics[height=4.cm]{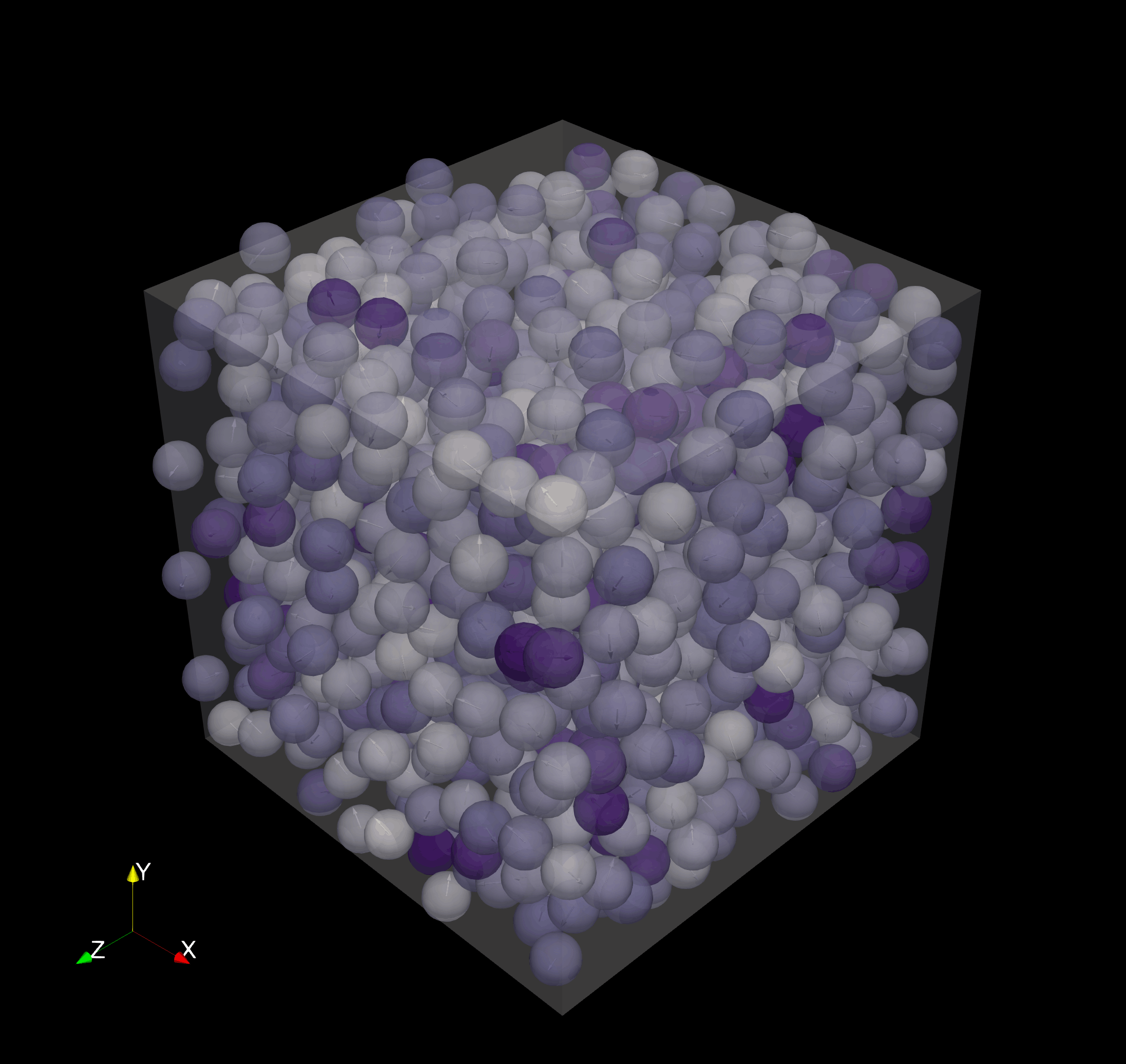} 
  \begin{picture}(0,0)
    \put(-10.,0.2){\textcolor{white}{$\phi=10\%$}} 
    \put(-5.65,0.2){\textcolor{white}{$\phi=20\%$}} 
    \put(-1.35,0.2){\textcolor{white}{$\phi=40\%$}} 
  \end{picture}
  \caption{Suspensions of 1024 shakers under shear ($B_2=0.2$, $\dot\gamma = 0.1$), colored by their instantaneous dissipation (darker particles are more dissipative), at various $\phi$.}
  \label{fig:snaps}
\end{figure*}

We perform large-scale hydrodynamic simulations of shaker or ABP suspensions using the active FSD method \citep{fiore2019fast, elfring2022active}.
Specifically, we simulate $N=1024$ particles in a periodic box at various volume fractions $\phi$, squirming coefficients $B_1$ and $B_2$, rotational diffusivity $d_r$ (thus prescribing $C_1$, up to an arbitrary rotation), and shear rates $\dot\gamma$; see \fig{fig:snaps} for illustrations and tables \ref{tab:shakers}--\ref{tab:ABPs} for the governing parameters.
Larger boxes with $N$ up to $8192$ at $\phi=10\%$ and $\dot\gamma = 10^{-3}$ are also simulated to examine the size dependence (see Appendix \ref{sec:N}).
Physically, there is only one `active' time scale in shaker suspensions, $\tau_a \equiv a/|B_2|$; whereas ABPs have two time scales, $\tau_s \equiv a/U_s$ due to self-propulsion, and $\tau_r \equiv \frac{1}{2} d_r^{-1}$ due to rotational diffusion. 
These time scales ($\tau_a$, or $\tau_s$ and $\tau_r$) compete with that of the shear, $\tau_{\dot\gamma} \equiv \dot\gamma^{-1}$, resulting in nonlinear dynamics as the P\'eclet number, Pe $\equiv \dot\gamma a^2/D_0$, varies.
Here, $D_0$ is the intrinsic diffusivity of the suspension without shear, which depends on $\tau_a$ and $\phi$ in the case of shakers, or $U_s$, $\tau_r$ and $\phi$ in the case of ABPs \citep{Ge_Brady_Elfring2025}.
For squirmers in general, all these parameters would affect $D_0$, resulting in a much larger phase space.
Theoretically, the interparticle force $\bm F_\text{ext}$ could also introduce a time scale, $\tau_e \equiv 6\pi\eta a^2/F_0$, with $\eta$ being the dynamic viscosity of the fluid.
However, $\bm F_\text{ext}$ is only activated when particles are very close and $\tau_e$ is fixed at a small value to model excluded-volume interactions; practically, we set $\kappa^{-1}/a=0.01$ and $\tau_e =0.1$. 
Once the particle velocities $\mathbf U$ are obtained, the particle positions and orientations are integrated forward in time using the standard Euler-Maruyama method with a timestep $\Delta t = 0.001$.
The choice of these parameters is consistent with our previous work \citep{Ge2022, Ge_Elfring_2025, Ge_Brady_Elfring2025}.
Finally, the FSD solver is implemented as a plugin to an open-source molecular dynamics package, called HOOMD-blue \citep{hoomd-blue}.
The implementation of our solver is also publicly available (see Data availability statement).

\begin{table*}
  \centering
  \setlength{\tabcolsep}{14pt} \renewcommand{\arraystretch}{1.2}
  \begin{tabular}{cccccccc}
    \hline
    $\phi$   &   $B_1$   &   $B_2$    &   $\tau_a$ & $\dot\gamma$  \\
    \hline
    10\%, 20\%, 30\%, 40\% & 0 & $\pm 0.05$ & 20 &  $[0, 3]$\\  
    10\%, 20\%, 30\%, 40\% & 0 & $\pm 0.10$ & 10 &  $[0, 3]$\\  
    10\%, 20\%, 30\%, 40\%& 0 & $\pm 0.20$ & 5 &  $[0, 3]$\\  
    5\%, 15\%, 25\%, 35\%, 45\% &  0 & $\pm 0.20$  &  5  &  0\\ 
    \hline
  \end{tabular}
  \caption{Parameters of the shaker simulations, where $\tau_a \equiv a/|B_2|$ and $a=1$.}
  \label{tab:shakers}
\end{table*}

\begin{table*}
  \centering
  \setlength{\tabcolsep}{14pt} \renewcommand{\arraystretch}{1.2}
  \begin{tabular}{cccccccc}
    \hline
    $\phi$   &   $B_1$   &  $B_2$ & $U_s$  & $\tau_r$ & $\ell_{p,0}$ & $\dot\gamma$ \\
    \hline
    10\% & 0.01 & 0  & 1/150  & 15  & 0.1    &  $[0, 3]$  \\  
    10\% & 0.01 & 0  & 1/150  & 45  & 0.3    &  $[0, 3]$  \\  
    10\% & 0.05 & 0  & 1/30   & 30   & 1.0    &  $[0, 3]$  \\  
    10\% & 0.10 & 0  & 1/15   & 45   & 3.0    &  $[0, 3]$  \\  
    10\% & 0.10 & 0  & 1/15   & 150  & 10.0 &  $[0, 3]$   \\  
    \hline
  \end{tabular}
  \caption{Parameters of the ABP simulations, where $U_s \equiv \frac{2}{3}B_1$ and $\ell_{p,0} \equiv U_s\tau_r$.}
  \label{tab:ABPs}
\end{table*}

\section{Results}
\label{sec:results}

\subsection{Dissipation}
\label{sec:edi}

\begin{figure*}
  \centering
  \includegraphics[height=6cm]{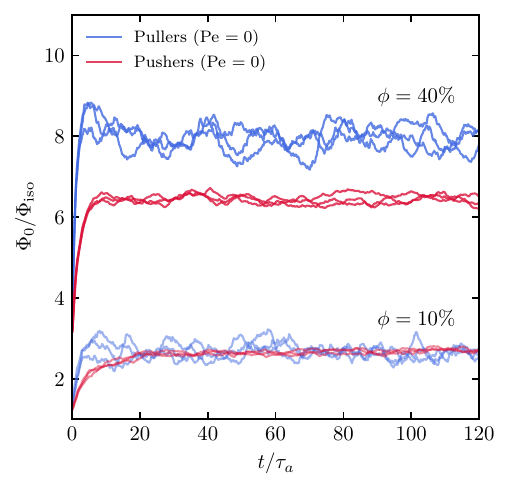}
  \includegraphics[height=6cm]{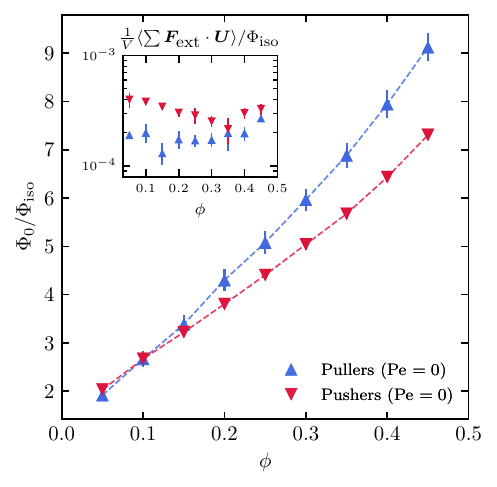}
  \begin{picture}(0,0)
    \put(-12.4, 6){(a)} \put(-6.3,6){(b)}
  \end{picture}
  \caption{Dissipation in suspensions of shakers without shear. 
  (a) Temporal evolutions of the internal dissipation, relative to that of isolated shakers at the same volume fraction $\phi$, for pulling and pushing shakers at $\phi=10\%$ and 40\%.
  Three random initial conditions are plotted in each case.
  (b) Average dissipation at steady state for both pullers and pushers, where the vertical bars represent one standard deviation (same in other similar plots).
    Inset shows the \xx{average power of the interparticle forces per unit volume for pullers and pushers at all $\phi$ and steady state, which is negligible relative to $\Phi_\text{iso}$ (see Appendix~\ref{sec:edi_int}).}}
  \label{fig:phi-edi0}
\end{figure*}
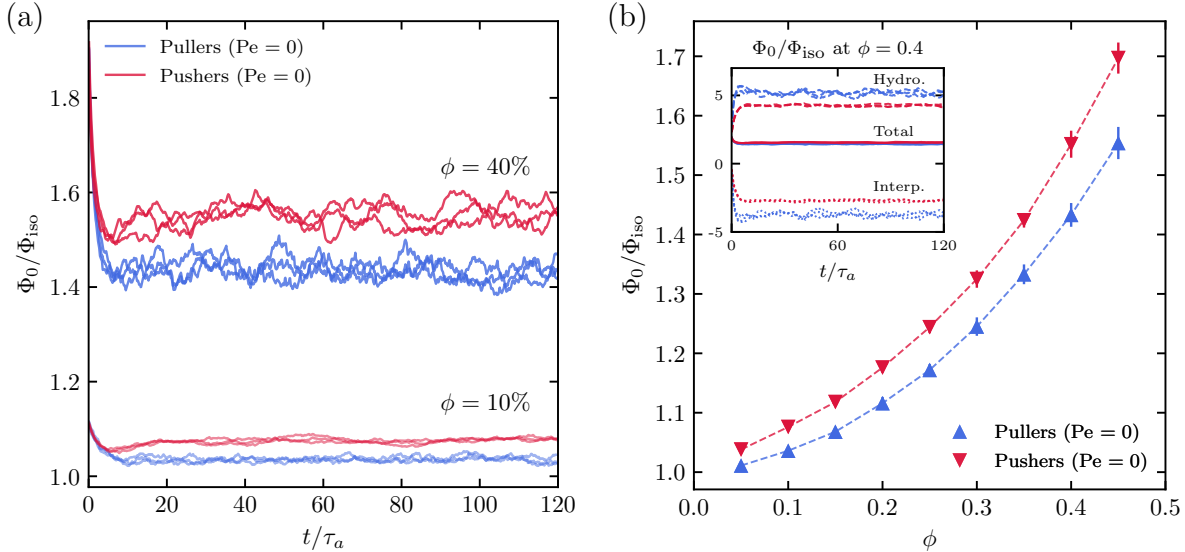

Now we examine the dissipation in non-dilute suspensions of shakers.
\Fig{fig:phi-edi0} shows $\Phi_0/\Phi_\text{iso}$ for pulling and pushing shakers at various volume fractions without external flow.
Here, $\Phi_\text{iso}$ is evaluated using Eq.~\eqref{eq:edi-iso} at the same $\phi$ that $\Phi_0$ is measured, as if the particles were \emph{not} interacting; thus, $\Phi_0/\Phi_\text{iso}$ quantifies the effect of particle interactions on the dissipation.
\Fig{fig:phi-edi0}(a) shows that the instantaneous dissipations fluctuate around certain mean values, implying that the suspension microstructures have been driven to their steady configurations.
At steady state, we observe $\Phi_0/\Phi_\text{iso} >1$ in all cases, suggesting that particle interactions \emph{enhance} dissipation.
\xx{This enhancement is hydrodynamic in nature, as the interparticle repulsion does no net work (its power, $\frac{1}{V}\sum \bm F_\text{ext}\bm\cdot\bm U$, becomes negligible relative to $\Phi_\text{iso}$, see \fig{fig:phi-edi0}(b) inset), and it affects $\Phi_0$ only through the hydrodynamic stresslet, \Eq{eq:S_H}, and the suspension microstructure.}
\xx{Furthermore, \fig{fig:phi-edi0}(b) shows that pullers tend to dissipate more energy than pushers, particularly at higher volume fractions ($\phi \gtrapprox 0.2$).}
Previously, we have shown that apolar active suspensions of shakers display activity-induced hydrodynamic diffusion, where pullers are more diffusive than pushers \citep{Ge_Elfring_2025}.
\xx{Interestingly, dissipation, also arising from particle activity, exhibits a similar trend.}

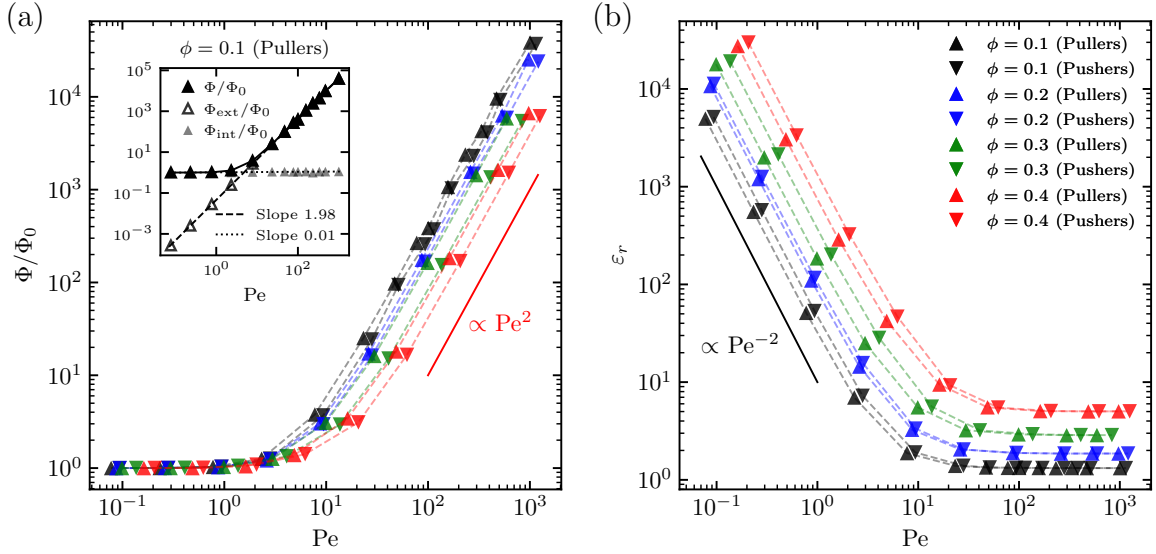
\begin{figure*}
  \centering
  \includegraphics[height=6cm]{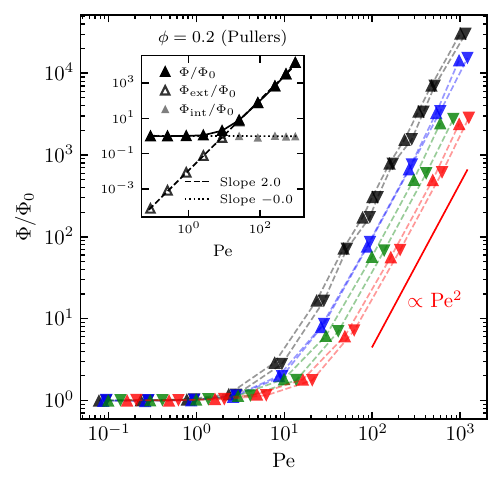}
  \includegraphics[height=6cm]{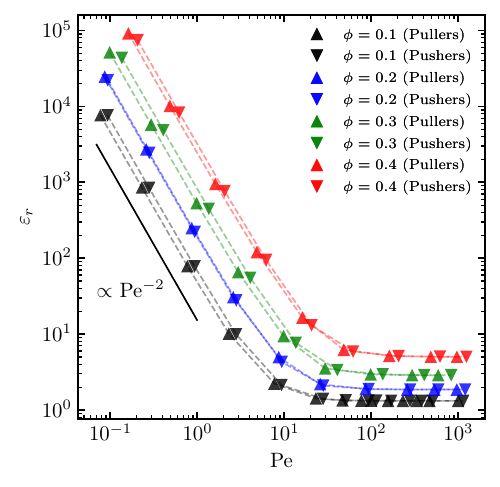}
  \begin{picture}(0,0)
    \put(-12.3,6){(a)} \put(-6.2,6){(b)}
  \end{picture}
  \caption{(a) Total dissipation under shear $\Phi$, relative to its value without shear $\Phi_0$, as a function of Pe at different volume fractions $\phi$ for suspensions of pullers and pushers; see panel (b) for the legends.
  Inset shows a decomposition of $\Phi$ into $\Phi_\text{ext}$ and $\Phi_\text{int}$ for pulling shakers at \xx{$\phi=20\%$}.
  (b) Relative dissipation $\varepsilon_r$ as a function of Pe at different $\phi$.}
  \label{fig:Pe-edi}
\end{figure*}

Next we examine the effect of shear on the dissipation in suspensions of shakers. 
\Fig{fig:Pe-edi}(a) shows the total dissipation $\Phi$, relative to the internal dissipation $\Phi_0$, for both pullers and pushers under various $\phi$ and Pe.
The most salient feature here is the two physical regimes depending on Pe. 
When Pe $<1$, we observe an activity-dominated dissipation, where all data collapse and $\Phi \approx \Phi_0$.
This is also the regime where $D \approx D_0$, where $D$
is the transverse diffusivity (in the $yz$-plane) under shear \citep{Ge_Brady_Elfring2025}.
Therefore, activity dominates both dynamically and mechanically when Pe $<1$.
Conversely, when Pe $\gg 1$, we observe $\Phi/\Phi_0 \sim$ Pe$^2$, a shear-driven scaling corresponding to constant relative dissipation, since constancy of $\varepsilon_r$ implies $\Phi /\Phi_0\sim \Phi_f/\Phi_\text{iso} \sim (\eta \dot\gamma^2)/(\eta\tau_a^{-2}) \sim (\dot\gamma \tau_a)^2 \sim$ Pe$^2$ (we have dropped the $\phi$-dependence in these scalings).
As shown in \fig{fig:Pe-edi}(b), $\varepsilon_r$ indeed becomes independent of Pe when Pe $\gtrapprox 100$, where the dynamical scaling $D/D_0 \sim$ Pe is also observed \citep{Ge_Brady_Elfring2025}.
Thus, shear dominates both dynamically and mechanically when Pe $\gtrapprox 100$.
As a side note, the second threshold, Pe$_s \approx 100$, results from the competition of two time scales: above Pe$_s$, even the \emph{longest} time scale due to shear is \emph{shorter} than the \emph{shortest} time scale due to activity \citep{Ge_Brady_Elfring2025}.
Pe$_s$ happens to be around 100 for the shakers we consider, but its value may change for other types of particles (\eg ABPs). 

Finally, the crossover from an activity- to shear-dominated regime can be understood by decomposing the total dissipation into its external and internal components.
The inset in \fig{fig:Pe-edi}(a) shows that $\Phi_\text{ext}/\Phi_0 \sim$ Pe$^2$, while $\Phi_\text{int}/\Phi_0 \approx 1$, for pullers at \xx{$\phi=20\%$} (similarly for pushers and at other $\phi$); both scalings remain nearly unchanged across the entire range of Pe considered.
These two curves cross at a certain Pe that increases with the volume fraction $\phi$ due to the elevated $\Phi_0$ at larger $\phi$; cf.~\fig{fig:phi-edi0}(b).
Consequently, $\Phi/\Phi_0$ reduces with $\phi$ at a fixed Pe in the shear-dominated regime, as shown in \fig{fig:Pe-edi}(a). 

\subsection{Rheology}
\label{sec:rheo}

The preceding results suggest that both hydrodynamic interactions and shear enhance the \emph{total} dissipation in shaker suspensions. 
In practice, it may be desirable to know a material's mechanical response to \emph{external} deformation explicitly, particularly for active suspensions known to display such unusual rheological signatures as the superfluid transition \citep{Saintillan2018}.
Therefore, in this section, we examine the rheological properties, characterized by the relative viscosity and normal stress differences, of our shaker and ABP suspensions.

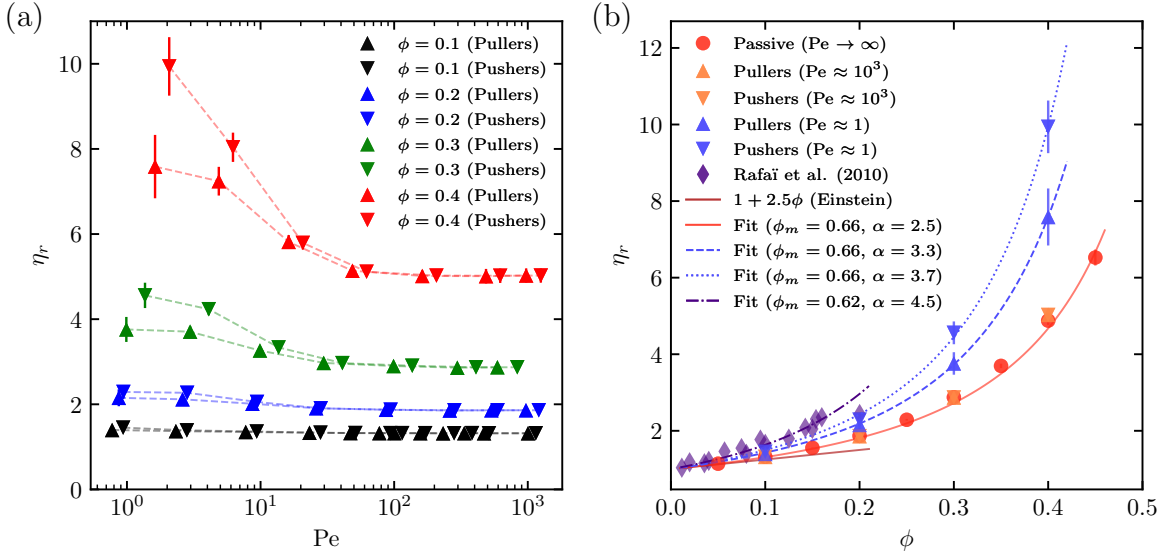
\begin{figure*}
  \centering
  \includegraphics[height=6cm]{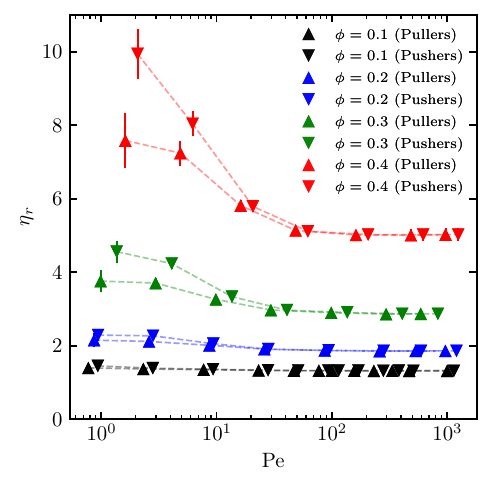}
  \includegraphics[height=6cm]{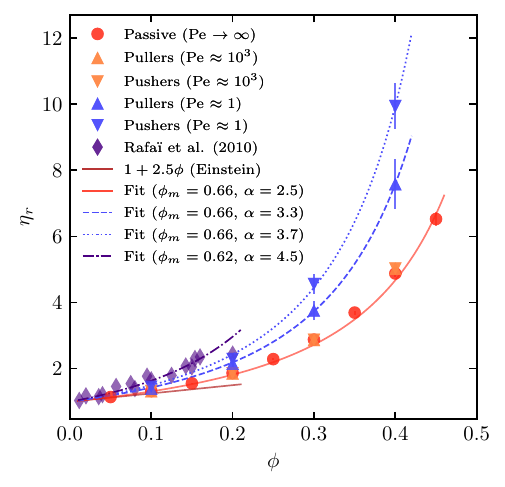}
  \begin{picture}(0,0)
    \put(-12.4,6){(a)} \put(-6.2,6){(b)}
  \end{picture}
  \caption{Relative viscosity $\eta_r$. 
  (a) Rate-dependence of $\eta_r$ at different volume fractions $\phi$ for suspensions of pulling or pushing shakers.
  (b) $\phi$-dependence of $\eta_r$ for pulling or pushing shakers at Pe $\approx 1$ and $10^3$, as well as for passive particles as Pe $\to \infty$ (by turning off activity).
  The experimental results of \citet{Rafai2010} at Pe $\approx 0.1$ are also plotted; cf.~Fig.~2b therein.
  Lines are the Einstein viscosity or fits to Eq.~\eqref{eq:kd}.}
  \label{fig:visc}
\end{figure*}

\begin{figure*}
  \centering
  \includegraphics[height=6cm]{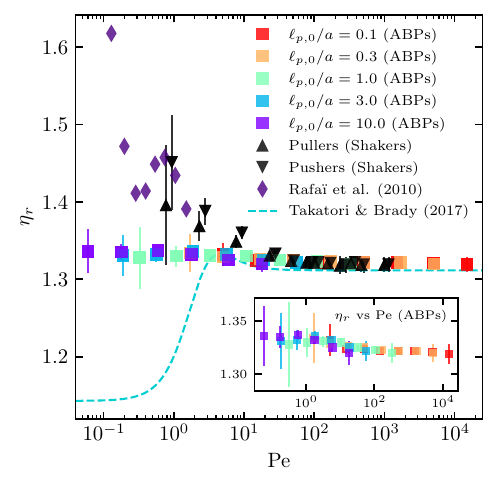}
  \includegraphics[height=6cm]{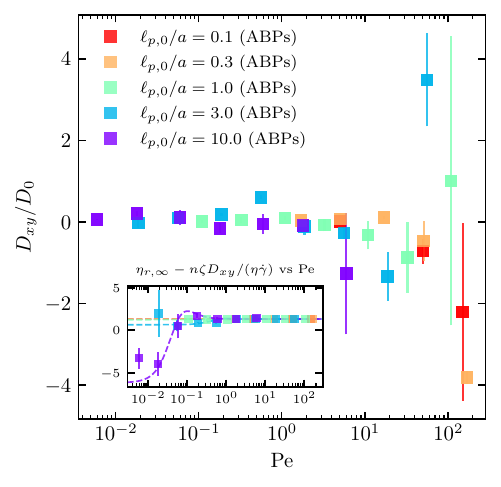}
  \begin{picture}(0,0)
    \put(-12.3,6){(a)} \put(-6.2,6){(b)}
  \end{picture}
  \caption{(a) Relative viscosity as a function of Pe for ABPs, shakers, and \emph{C.~reinhardtii} (extracted from Fig.~2(a) in \citet{Rafai2010}, based on $a=5$ $\mu$m and $D=995$ $\mu$m$^2/s$) at 10\% volume fraction.
  The dashed line is the theoretical prediction of \citet{Takatori2017} at $\ell_{p,0}/a=1.5$.
  Inset shows the ABP viscosity as a function of Pe.
  (b) The normalized cross-diffusivity, $D_{xy}/D_{0}$, as a function of Pe for ABPs at $\phi=10\%$ (data with large uncertainties are removed).
  Inset shows the theoretical relative viscosity predicted from the particle dynamics, where markers correspond to direct application of the swim stress-diffusivity relation using the fitted $D_{xy}$, and lines are calculations based on Eq.~\eqref{eq:Takatori} (same color coding as in the main figure).
  }
  \label{fig:visc-phi10}
\end{figure*}

\Fig{fig:visc}(a) shows the relative viscosity $\eta_r$ as a function of Pe for suspensions of pulling or pushing shakers at various volume fractions $\phi$ (results at Pe $\ll 1$ have large fluctuations and thus are not included; see Appendix \ref{sec:N} for details).
Overall, we observe a \emph{shear-thinning} rheology regardless of the particle type, but more so for pushers and at higher $\phi$; see \fig{fig:visc-phi10}(a) for a closer look of the results at $\phi=10\%$.
Furthermore, the range of Pe where shear-thinning occurs seems to coincide with the region of nonmonotonic diffusion, $1 \lessapprox$ Pe $\lessapprox 100$, consistent with our observation of the total dissipation, implying that all these behaviors may share a common origin.
In the literature, it is often reported that pullers (spherical or not) shear thin \citep{Rafai2010, Mussler2013, saintillan2010dilute}, while pushers (typically elongated) shear thicken \citep{Gachelin2013, lopez2015turning, Martinez2020}, which may give the impression that the rate-dependence mainly depends on the swimming mechanism.
Here, our results provide a clear counter-example, highlighting the importance of particle shape in the rheology of active suspensions.

The relative viscosity can also be plotted as a function of the volume fraction to better examine the existence (or absence) of the superfluid transition.
\Fig{fig:visc}(b) shows $\eta_r$ for pulling or pushing shakers at Pe $\approx 1$ and $10^3$, corresponding to the activity- and shear-dominated regimes, respectively, as well as for passive particles as activity is turned off (thus Pe $\to \infty$).
Unlike bacterial suspensions where $\eta_r$ reduces with $\phi$ at sufficiently low shear rates (or Pe) \citep{Martinez2020}, our shakers always exhibit enhanced viscosity as $\phi$ increases.
This resembles the behavior of passive suspensions, for which the relative viscosity can be described by various empirical relations, such as that of \citet{krieger1959mechanism}
\begin{equation} \label{eq:kd}
  \eta_r = \bigg( 1- \frac{\phi}{\phi_m} \bigg)^{-\alpha \phi_m},
\end{equation}
where $\phi_m$ is the maximal packing fraction, and $\alpha$ is the `intrinsic viscosity.'
Setting $\alpha=2.5$ recovers Einstein's viscosity law as $\phi \to 0$.
Although Eq.~\eqref{eq:kd} can be inaccurate at higher $\phi$ \citep{guazzelli_pouliquen_2018}, we use it to fit our data nonetheless to compare directly with a previous experiment on algal (\emph{C.~reinhardtii}) suspensions \citep{Rafai2010}.
Specifically, we first fit $\phi_m \approx 0.66$ from the passive suspensions using $\alpha=2.5$, then keep the same $\phi_m$ to fit $\alpha$ for the active suspensions, obtaining $\alpha =$ 3.3 (pullers) and 3.7 (pushers) at Pe $\approx 1$; see \fig{fig:visc}(b). 
Similarly, \citet{Rafai2010} reported $\phi_m=0.62$ and $\alpha=2.5 \pm 0.1$ for dead algae (equivalent to passive particles, Pe $\gg 1$), and $\alpha=4.5 \pm 0.2$ for live algae (we estimate their Pe $=\dot\gamma a^2/D \approx 0.1$, based on the shear rate $\dot\gamma=5$ s$^{-1}$, cell radius $a=5$ $\mu$m, and diffusivity $D=995$ $\mu$m$^2/$s).
This qualitative agreement indicates that the enhanced viscosity may be related to particle activity, as opposed to the possible bottom-heaviness or non-spherical shape of the algae \citep{Rafai2010}.
Indeed, later experiments using the same algae by \citet{Mussler2013} also suggested that the last two possibilities might be unlikely.

One major difference between our simulation and the experiments of \citet{Rafai2010} and \citet{Mussler2013} is that our shakers are individually immotile, while the algae \emph{C.~reinhardtii} are highly motile.\footnote{Another potentially important difference is that \emph{C.~reinhardtii} is not simply a puller. As shown by \citet{Drescher2010}, the averaged flow field around the alga can be approximated by three Stokeslets, which only resembles a puller-stresslet at distances much greater than the particle size.}
This can be clearly seen in the persistent length of their swimming motion: an isolated shaker cannot swim, while a \emph{C.~reinhardtii} can swim 28 body lengths before turning around ($\ell_{p,0}/a \equiv U_\text{s} \tau_r/a \approx 40 \times 3.5 / 5 = 28$).
Therefore, to separate the effects of motility from activity, we simulate self-propelling but non-hydrodynamically interacting neutral squirmers, or ABPs, with different intrinsic persistence lengths; cf.~table \ref{tab:ABPs}. 
\Fig{fig:visc-phi10}(a) shows the corresponding relative viscosity $\eta_r$ as a function of Pe for the ABPs, together with shakers and \emph{C.~reinhardtii}, at $\phi=10\%$.
Remarkably, the relative viscosities of ABPs at different $\ell_{p,0}$ collapse onto a single curve, displaying a weak shear-thinning rheology; see \fig{fig:visc-phi10}(a) inset (the data at Pe $\ll 0.1$ have large fluctuations and are thus not included; see Appendix \ref{sec:N}).
The much weaker rate-dependence of the relative viscosity of ABPs confirms that the enhanced viscosity at low Pe seen in the algal experiments \citep{Rafai2010, Mussler2013} may be mainly due to particle activity, as opposed to motility.
We will discuss the microstructral origin of the observed rheology further in \S \ref{sec:struct}.

The rheological behavior of ABPs casts doubt on a recent theoretical model relating the stress generation and diffusive motion of self-propelled particles \citep{Takatori2014, Takatori2017}.
In this model, a `swim stress,' $\mathsfb\Sigma^\text{swim}$, was related to the diffusion tensor, $\mathsfbi D^\text{swim}$, due to swimming as
\begin{align}\label{eq:swim-stress}
  \mathsfb\Sigma^\text{swim} = -n\zeta \mathsfbi D^\text{swim}, 
\end{align}    
where $n$ is the particle number density, $\zeta$ is the hydrodynamic drag coefficient ($\zeta=6\pi\eta a$ for a rigid sphere in a Newtonian fluid), and $\mathsfbi D^\text{swim} = \frac{1}{3}U_\text{s}^2\tau_r \mathsfbi I$.
Assuming $\mathsfb\Sigma^\text{swim}$ to be the only active stress component apart from its passive counterpart, \citet{Takatori2017} derived an expression for the relative viscosity of a dilute suspension of ABPs (cf.~\Eq{eq:visc_sq}),
\begin{align}\label{eq:Takatori}
  \eta_r = \eta_{r,\infty} - 0.75 \phi \ell_{p,0}^2 \frac{1-(\text{Pe}^\prime/4)^2}{[1+(\text{Pe}^\prime/4)^2]^2},
\end{align} 
where $\text{Pe}^\prime \equiv 2\dot\gamma \tau_r =\frac{2}{3} (\ell_{p,0}/a)^2$Pe, and $\eta_{r,\infty}$ is the relative viscosity as Pe$^\prime \to\infty$.
Here, we have expressed Eq.~\eqref{eq:Takatori} in our notation and replaced the Einstein viscosity, $\eta_{r,\infty}=1+2.5\phi$, in the original Eq.~(5) in \citet{Takatori2017} with Eq.~\eqref{eq:kd} ($\phi_m=0.66$ and $\alpha=2.5$), as the latter fits our data better; see \fig{fig:visc}(b).
The dashed line in \fig{fig:visc-phi10}(b) shows the theoretical prediction at $\phi=0.1$ and $\ell_{p,0}/a=1.5$, which displays a shear thickening behavior at Pe $\lessapprox 4$ \emph{not} seen in our simulations.
Further increasing the persistence length would lead to larger viscosity reductions and even \emph{negative} viscosities (see figure 4 in \citet{Takatori2017}), again at odds with both numerical and experimental observations \citep{Rafai2010, Mussler2013}.
Therefore, the existing results do not directly support the theory.

To understand the potential reason for such discrepancy, we compute the $xy$ component of the diffusion tensor to calculate the theoretically predicted shear stress, since the theory asserts $\Sigma^\text{swim}_{xy} = -n\zeta D^\text{swim}_{xy}$. 
This requires knowledge of the mean square displacements $\langle \Delta y(t)^2 \rangle$ and covariance $\langle \Delta x(t)\Delta y(t) \rangle$, which exhibit the following limiting behaviors as $t \to \infty$ \citep{sierou2004shear}, 
\begin{align} \label{eq:msd}
  \big\langle \Delta y(t)^2 \big\rangle & \to 2D_{yy}t + C_{yy}, \\
  \big\langle \Delta x(t) \Delta y(t) \big\rangle & \to \dot\gamma D_{yy} t^2 + (2D_{xy} + \dot\gamma C_{yy}) t + C_{xy},
\end{align}
where $D_{ij}$ are the diffusion coefficients, $C_{ij}$ are the integration constants, and $\langle \cdot \rangle$ averages over all particles and reference time.
\Fig{fig:visc-phi10}(b) shows the normalized cross-diffusivity, $D_{xy}/D_{0}$, fitted from our ABP simulations (data with large uncertainties are removed).
The results do not display any clear trend due to the noisy data; however, the relative viscosity calculated from Eq.~\eqref{eq:Takatori} appears to agree with direct application of Eq.~\eqref{eq:swim-stress}, $\eta_{r,\infty} - n\zeta D_{xy}/(\eta \dot\gamma)$; see the inset in \fig{fig:visc-phi10}(b).  
Therefore, the particle dynamics is captured correctly by the theory (at least up to $\phi=10\%$), but the relation between diffusion and stress generation due to self-propulsion, Eq.~\eqref{eq:swim-stress}, may be untenable. 
In other words, the particle stress $\mathsfb\Sigma_p$ might not be represented by $-n\zeta \mathsfb D^\text{swim}$ for ABPs.

\begin{figure*}
  \centering
  \includegraphics[height=6cm]{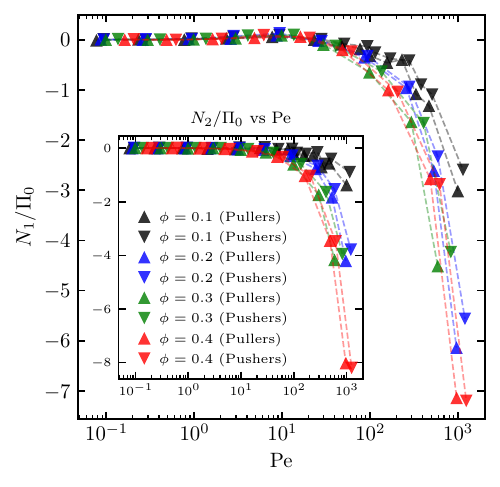}
  \includegraphics[height=6cm]{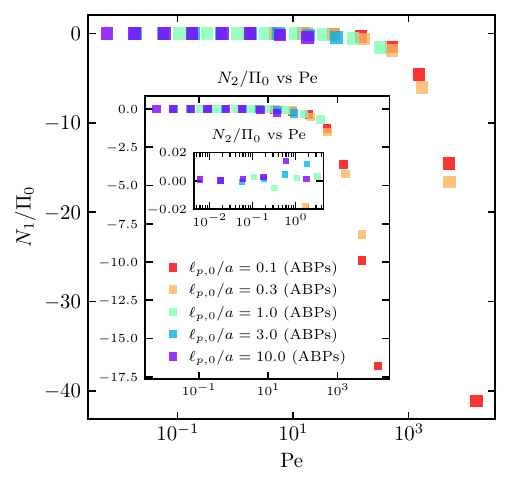}
  \begin{picture}(0,0)
    \put(-12.4,6){(a)} \put(-6.3,6){(b)}
  \end{picture}
  \caption{Rate dependence of the NSDs, normalized by the suspension pressure without shear, $N_1/\Pi_0$ (main figures) and $N_2/\Pi_0$ (insets).
  (a) Pulling and pushing shakers at different volume fractions.
  (b) ABPs at different persistence lengths; the smaller inset shows $N_2/\Pi_0$ at smaller Pe.}
  \label{fig:nsd}
\end{figure*}
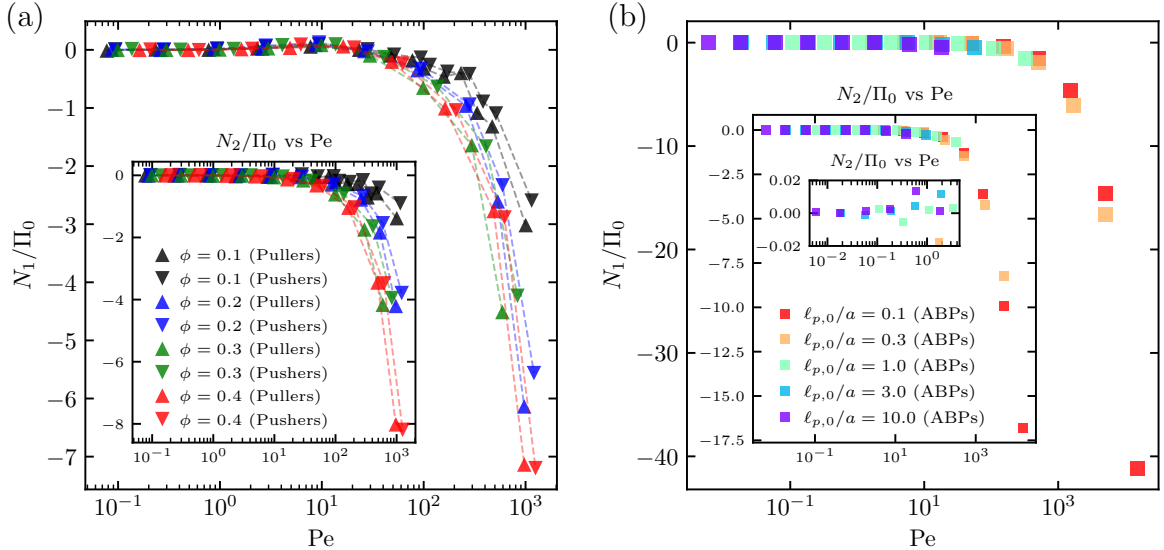

Finally, we examine the normal stress differences (NSDs), $N_1$ and $N_2$, defined as
\begin{align}\label{eq:nsd}
  N_1 \equiv \langle \Sigma_{p,xx} \rangle - \langle\Sigma_{p,yy} \rangle, \quad 
  N_2 \equiv \langle \Sigma_{p,yy} \rangle - \langle\Sigma_{p,zz} \rangle,  
\end{align} 
where $\langle\Sigma_{p,ii} \rangle$ are the diagonal components of the particle stress tensor, given by Eq.~\eqref{eq:Sigma_p}.
\Fig{fig:nsd} shows the NSDs normalized by the particle pressure at Pe $=0$, $\Pi_0 \equiv -\text{tr}(\mathsfb\Sigma_{p})/3$, for shakers and ABPs as functions of Pe.
Despite the extensive ranges of parameters considered, all data display the same general behavior, with  $N_1/\Pi_0 \approx 0$ and $N_2/\Pi_0 \approx 0$ at Pe $\lessapprox 10$, and $N_1/\Pi_0 < 0$ and $N_2/\Pi_0 < 0$ otherwise.
Negligible NSDs at small Pe suggests that the microstructure is isotropic, which is expected for suspensions of shakers or self-propelled particles without alignment interactions \citep{Vicsek1995, Simha_Ramaswamy2002, Ge_Elfring_2025}.
Conversely, the negative NSDs indicate a loss of isotropy, a behavior often found in passive suspensions under shear (Pe $\to \infty$) \citep{guazzelli_pouliquen_2018}.
In that case, non-hydrodynamic interactions or particle surface roughness result in an asymmetric microstructure, characterized by more particles in the compression quadrant of the shear plane \citep{morris2009review}.
Both these limiting behaviors are verified in our suspensions, as we will show in the next section.

Before we proceed, we may further scrutinize the swim stress-diffusion relation \citep{Takatori2014, Takatori2017} by comparing the theoretical predictions with our NSD data.
Eq.~\eqref{eq:swim-stress} implies $\Sigma_{p,ii} = -n\zeta D_{ii}$ and $\Pi_0 = n\zeta D_{0}$, thus $N_2/\Pi_0 = (D_{zz}-D_{yy})/D_{0}$.
For ABPs with $\ell_{p,0}/a=10$ at Pe $\approx 1$, our previous simulations showed that $D_{yy}/D_{0} \approx 0.04$ and $D_{zz}/D_{0} \approx 1$ \citep[see figure 3 in][]{Ge_Brady_Elfring2025}, yielding $N_2/\Pi_0 \approx 0.96$.
The inset in \fig{fig:nsd}(b) shows a much smaller value at Pe $\approx 1$, $N_2/\Pi_0 \approx 0.02$, not supporting the theory.

\subsection{Microstructure}
\label{sec:struct}

\begin{figure*}
  \centering
  \includegraphics[height=6cm]{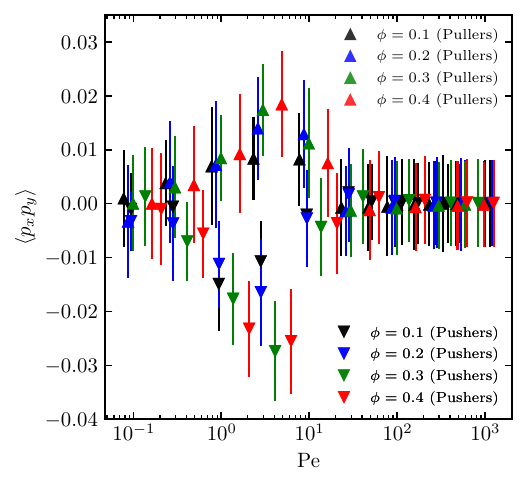}
  \includegraphics[height=6cm]{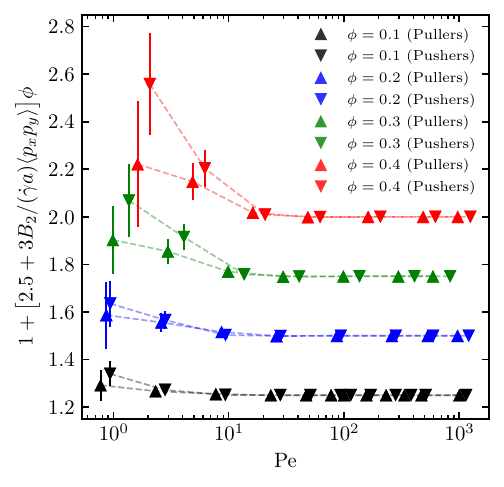}
  \begin{picture}(0,0)
    \put(-12.6,6){(a)} \put(-6.2,6){(b)}
  \end{picture}
  \caption{(a) Global orientation correlation, $\langle p_x p_y \rangle$, in shaker suspensions. 
  (b) The predicted relative viscosity based on \Eq{eq:visc_sq}, assuming no particle interactions.}
  \label{fig:pxpy}
\end{figure*}
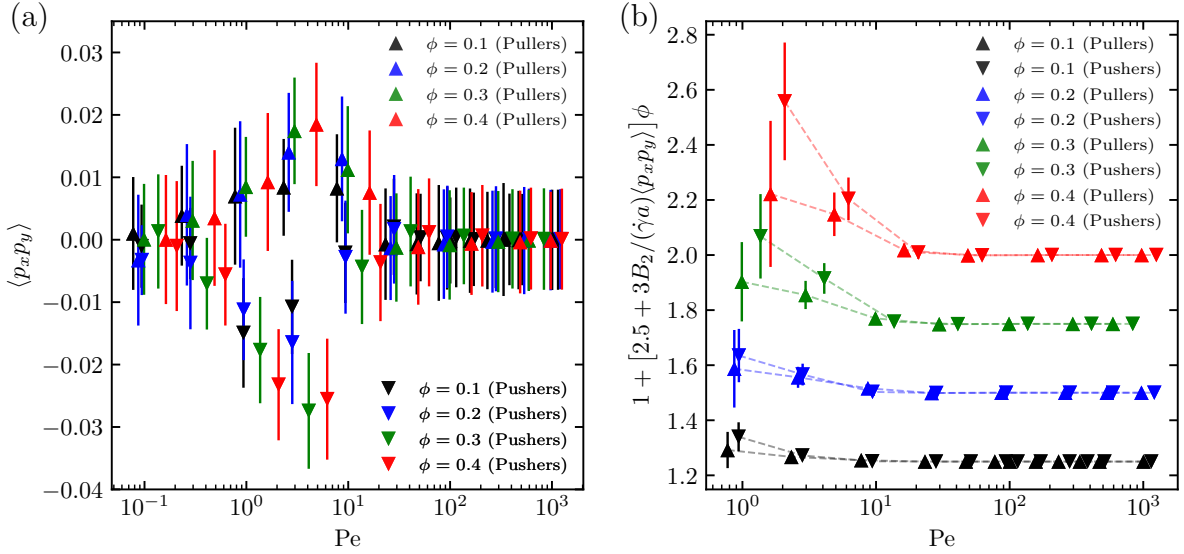

To gain a mechanistic understanding of the rheological responses, we analyze the suspension microstructure in detail in the remainder of the paper.
\Eq{eq:visc_sq} suggests that viscosity variations with Pe may arise if $B_2 \langle p_x p_y \rangle \ne 0$. 
In a dilute suspension of spherical active particles, no orientation correlation is expected because each particle simply rotates under shear with angular speed $\dot\gamma/2$.
In non-dilute suspensions, however, the particle orientations may become correlated due to hydrodynamic interactions.
\Fig{fig:pxpy}(a) shows $\langle p_x p_y \rangle$ for shaker suspensions under all $\phi$ and Pe considered. 
Indeed, we observe anisotropic averaged orientations at $1 \lessapprox$ Pe $\lessapprox 10$,  with $\langle p_x p_y \rangle >0$ for pullers and $\langle p_x p_y \rangle <0$ for pushers.
Substituting the corresponding values into \Eq{eq:visc_sq} yields a shear-thinning rheology in qualitative agreement with our previous observations; cf.~figures \ref{fig:visc}(a) and \ref{fig:pxpy}(b).
This simple calculation thus illustrates the effectiveness of using \Eq{eq:visc_sq} to approximate $\eta_r$, as well as the (perhaps) counter-intuitive effects of hydrodynamic interactions on $\langle p_x p_y \rangle$, in suspensions of shakers.

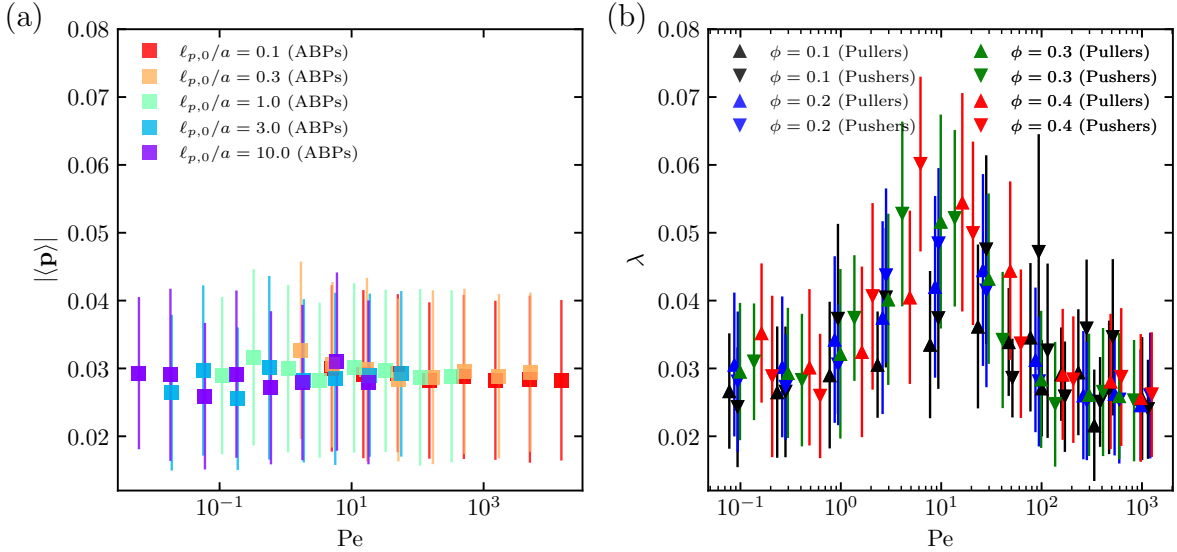
\begin{figure*}
  \centering
  \includegraphics[height=6cm]{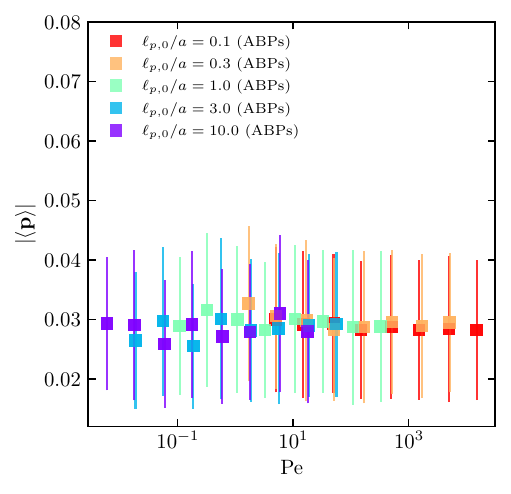}
  \includegraphics[height=6cm]{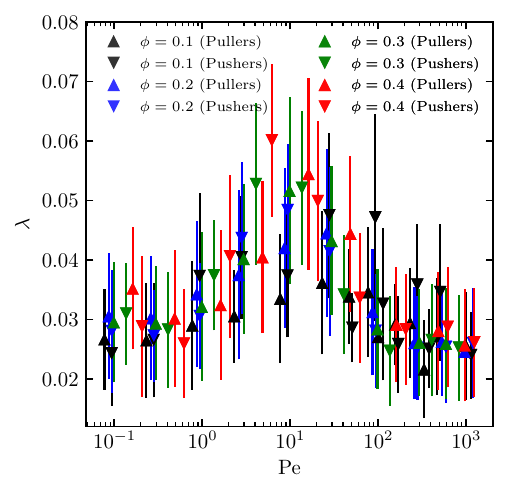}
  \begin{picture}(0,0)
    \put(-12.4,6){(a)} \put(-6.3,6){(b)}
  \end{picture}
  \caption{Global orientational orders.
  (a) Polar order of ABPs at various intrinsic persistence lengths and $\phi=$ 10\%.
  (b) Nematic order of pulling and pushing shakers at various $\phi$.
  }
  \label{fig:ori}
\end{figure*}

To characterize the suspension microstructure more generally, we examine the global orientational orders. 
For ABPs, the magnitude of the average direction of all particles, $|\langle \bm p \rangle|$, indicates the \emph{polar} order of the suspension.
\Fig{fig:ori}(a) shows that $|\langle \bm p \rangle|$ remains small ($\approx 1/\sqrt{N}$) for ABPs at any Pe, regardless of the intrinsic persistence length, suggesting that shear alone does not promote polar order.\footnote{Inspection of the orientation distributions shows a slight tendency for very persistent ABPs (e.g., $\ell_{p,0}/a=10$) to orient perpendicular to the shear plane at large Pe, possibly due to shear-induced collision. However, such tendency is nearly negligible as \fig{fig:ori}(a) indicates.}
For shakers, we can define a symmetric and traceless tensor, $\mathsfbi Q \equiv (3\langle \bm {pp} \rangle - \mathsfbi I)/2$, whose largest eigenvalue, $\lambda$, measures the \emph{nematic} order: $\lambda=0$ if the suspension is isotropic; $\lambda=1$ if it is maximally nematic \citep{Doi2013}.
\Fig{fig:ori}(b) shows that shakers have nearly isotropic orientations at small and large Pe, regardless of $\phi$; however, there is an \emph{enhanced} nematic order at 1 $\lessapprox$ Pe $\lessapprox$ 100, which appears to increase with $\phi$ for both pullers and pushers.
This behavior is similar to that of $\langle p_x p_y \rangle$ as both indicate nematic order, though the ranges of Pe where the deviations occur differ (explained later).
Theoretically, we expect $\lambda$ to vanish at small Pe because long-range nematic order is unstable when Pe $=0$ \citep{Simha_Ramaswamy2002}; this is consistent with our observations of the NSDs, see \fig{fig:nsd}.
At large Pe, the particles are effectively passive spheres, whose orientations are randomized by shear-induced hydrodynamic interactions and have no impact on the translational dynamics.
It is remarkable that the interplay of activity and shear leads to a \emph{nonmonotonic} nematic order which peaks at Pe $\approx 10$.

\begin{figure*}
  \centering
  \includegraphics[width=\columnwidth]{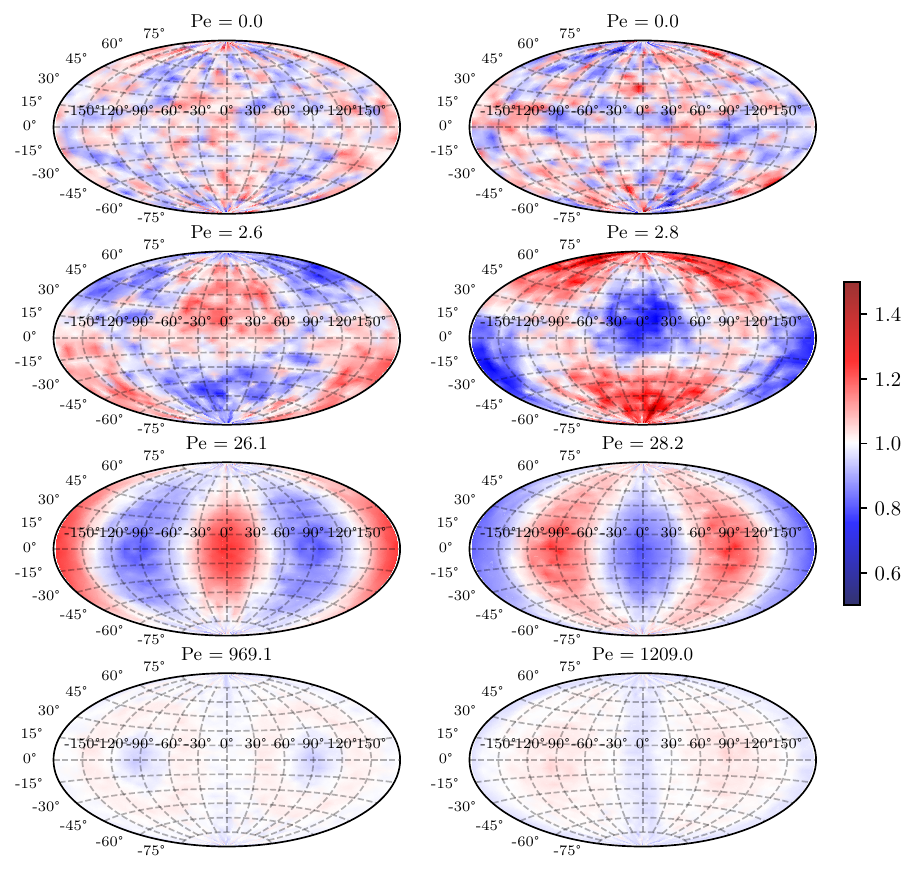}
  \begin{picture}(0,0)
  \end{picture}
  \caption{Orientation distributions for pullers (left column) and pushers (right column) under increasing Pe at $\phi=20\%$.
  The maps are oriented such that the flow direction ($x$) is toward the reader from the center, the velocity gradient direction ($y$) is from the south to north poles, and the vorticity direction ($-z$) is eastward from 90$^\circ$ on the equator.
  Color indicates the relative probability of the orientation, $P(\bm p)$.}
  \label{fig:map-ori}
\end{figure*}

To visualize how shakers orient as Pe increases, we define a distribution function, 
\begin{align}\label{eq:P(p)}
   P (\bm p) \equiv \xx{2\pi} \Big\langle \sum_{i=1}^N \big[\delta(\bm p - \bm p_i)+\delta(\bm p +\bm p_i) \big]\Big\rangle,
\end{align}
where $\delta$ is the Dirac delta \xx{on the unit sphere}, and $\langle \cdot \rangle$ averages over all particles and time at steady state.
By definition, $P(\bm p) = P(-\bm p)$ as $\pm\bm p$ are equivalent for shakers, and $P(\bm p)=1$ corresponds to a uniform distribution.
\Fig{fig:map-ori} shows $P(\bm p)$ for both pullers and pushers at a few representative Pe and $\phi= 20\%$ (the results at other volume fractions are similar).
Clearly, there are \emph{flow-induced alignments} at intermediate Pe, though the preferred orientation varies as Pe increases.
At 1 $\lessapprox$ Pe $\lessapprox$ 10, pullers are more likely to orient in the extension quadrant of the shear plane, thus $\langle p_x p_y \rangle >0$; while pushers in the compression quadrant, $\langle p_x p_y \rangle <0$.
At 10 $\lessapprox$ Pe $\lessapprox$ 100, pullers become nearly parallel to the flow, while pushers are perpendicular to it; in both cases, $\langle p_x p_y \rangle \approx 0$.
The nematic ordering is apparent across the entire range of 1 $\lessapprox$ Pe $\lessapprox$ 100.
Similar alignment has been observed experimentally in spherical puller-type algal suspensions \citep{Rafai2010}.
Our results thus generalize their finding and provide an alternative explanation for the alignment due to activity under hydrodynamic interactions.

\begin{figure*}
  \centering
  \includegraphics[height=6cm]{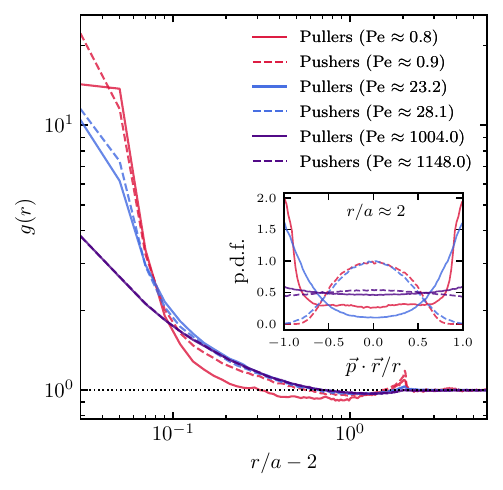}
  \includegraphics[height=6cm]{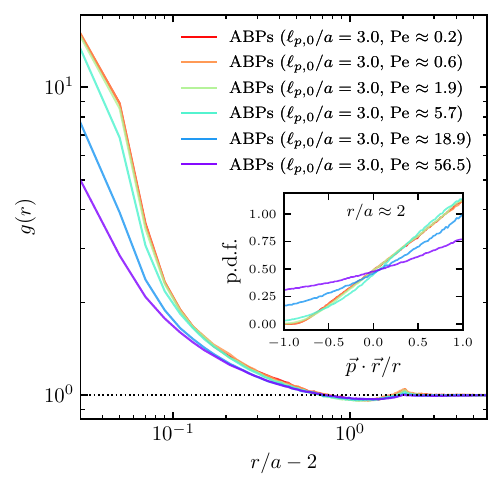}
  \begin{picture}(0,0)
    \put(-12.4,6){(a)} \put(-6.3,6){(b)}
  \end{picture}
  \caption{\xx{Radial distribution function, $g(r)$,} for suspensions of pulling or pushing shakers (a) and ABPs (b) at 10\% volume fraction.
  \xx{The insets show the probability density functions (p.d.f.) of} $\bm p \cdot \bm r/r$ near contact.}
  \label{fig:g(r)}
\end{figure*}

To quantify the importance of hydrodynamic interactions, as well as to examine the \emph{local} order, we calculate the pair correlation function, 
\begin{equation} \label{eq:g(r)}
  g({\bm r}) = \xx{\frac{1}{n}} \Big\langle \sum_{i \ne j}^N \delta(\bm r-\bm r_{ij}) \Big\rangle,
\end{equation}
\xx{where $\bm r_{ij}$ is the separation vector between particles $i$ and $j$, and $\langle \cdot \rangle$ averages over all particles and time at steady state.
Its angular average gives the radial distribution, $g(r) = \langle \sum_{i \ne j}^N \delta(r - r_{ij}) \rangle /(4\pi r^2 n)$, with $r_{ij} = |\bm r_{ij}|$.}
For suspensions of shakers without shear, our previous simulations showed that the radial distributions, $g(r)$, oscillate with decaying peaks at the first few particle layers: typically at $r/a \approx 2$, 4, 5.5, and so on.
Meanwhile, the angular distributions in the first particle layer are anisotropic: pullers tend to find their nearest neighbors along the particle axis (\ie~$|{\bm p}\cdot{\bm r}|/r \approx 1$), whereas pushers tend to find them in the plane perpendicular to ${\bm p}$ (\ie~${\bm p}\cdot{\bm r}/r \approx 0$) \citep{Ge_Elfring_2025}.
\Fig{fig:g(r)}(a) shows that these effects are \emph{weakened} at larger Pe and  $\phi= 10\%$ (similarly at other volume fractions), as both the radial and angular distributions flatten under increasing Pe.
Similar behaviors are also generally observed in suspensions of ABPs, though the angular distributions are asymmetric due to self-propulsion, as shown in \fig{fig:g(r)}(b) for the case of $\ell_{p,0}/a=3$. 
In \S \ref{sec:edi}, we have seen that hydrodynamic interactions, which are dominated by particles in close contact, enhance dissipation.
Therefore, the reduced $g(r\approx 2a)$ implies the reduced dissipation, corroborating the reducing nematic order as Pe increases from about 10 (cf.~figure \ref{fig:ori}b),  consistent with the shear-thinning behaviors observed in \figs{fig:Pe-edi}(b) and \ref{fig:visc}(a).

\begin{figure*}
  \centering
  \includegraphics[width=\columnwidth]{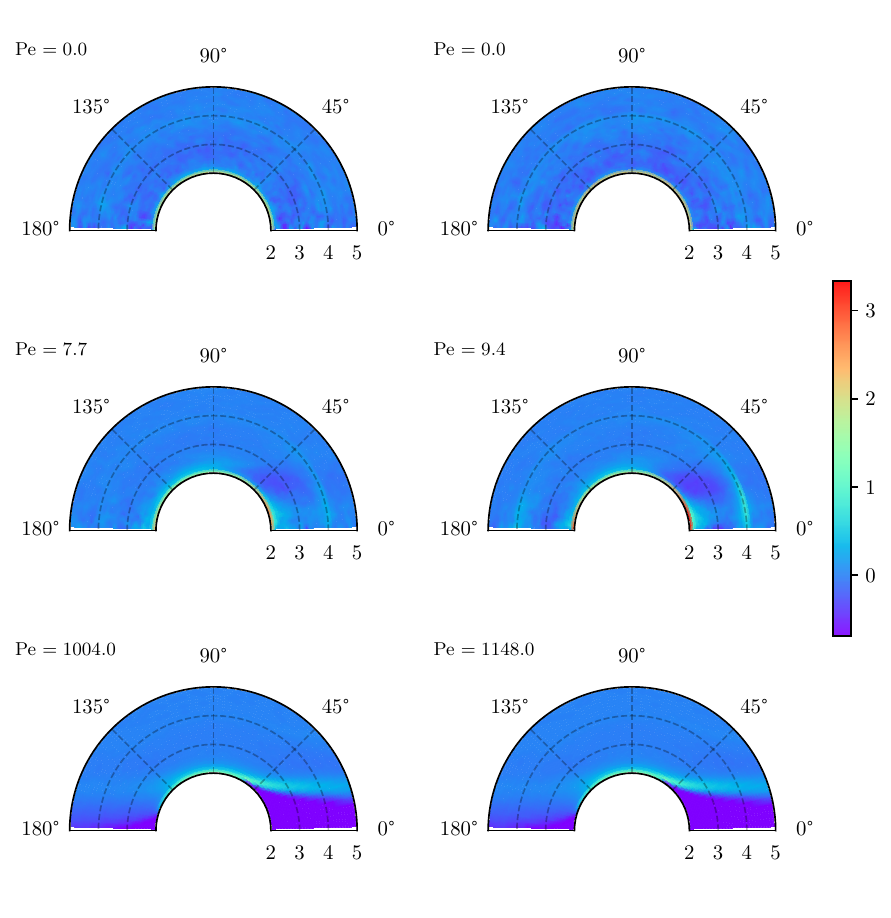} 
  \begin{picture}(0,0)
  \end{picture}
  \caption{Pair correlation functions projected onto the shear plane for pullers (left column) and pushers (right column) at $\phi=10\%$, where the flow direction ($x$) is horizontal (0$^\circ$) and the velocity gradient direction ($y$) is vertical (90$^\circ$).
  Color denotes $\ln[g(\bm r)]$.}
  \label{fig:adf-shear}
\end{figure*}

\begin{figure*}
  \centering
  \includegraphics[width=\columnwidth]{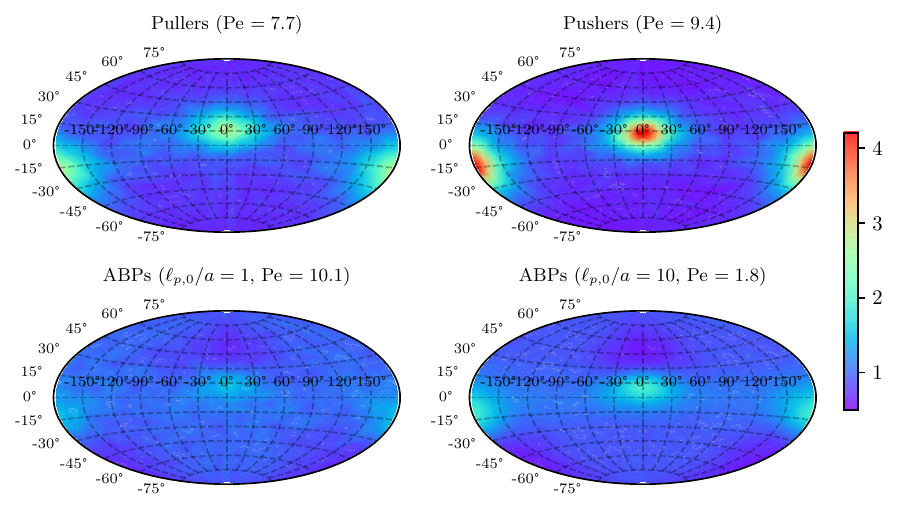} 
  \begin{picture}(0,0)
  \end{picture}
  \caption{Angular distributions of the nearby particles ($r/a \approx 2$) for shakers and ABPs at $\phi=10\%$.
  The maps are oriented in the same way as \fig{fig:map-ori}.
  Color indicates $P(\bm r/r)$, \ie~the relative probability of finding a neighbor at position $\bm r$; cf.~\Eq{eq:P(p)}.}
  \label{fig:map-ori-r}
\end{figure*}

Finally, we can plot the pair correlation functions projected onto the shear plane, $g(r,\theta)$, where $\cos\theta=\bm e_x \cdot \bm r/r$ and $\bm e_x$ is the flow direction.
\Fig{fig:adf-shear} shows $g(r,\theta)$ for both pullers and pushers at $\phi=10\%$ under increasing Pe.
As expected, we observe an isotropic distribution at Pe $=0$ due to activity-induced hydrodynamic interactions, and an anisotropic structure characterized by an accumulation of particles at contact in the compression and depletion of particles in the tension quadrants at large Pe \citep{morris2009review}.
Both of these are consistent with our observations of the NSDs, cf.~\fig{fig:nsd}.
Most interestingly, at Pe $\approx 10$ where neither activity nor shear dominates, the microstructure displays an anisotropy with more particle pairs in the tension quadrant roughly aligned with the flow, \ie~$0^\circ < \theta < 45^\circ$ (note that the colorbar is logarithmic, thus the probability is large).
The distribution at $r/a \approx 2$ is also plotted in \fig{fig:map-ori-r}, which clearly highlights the anisotropy.
Furthermore, the tendency is so strong that it could even be seen in the second particle layer at $r/a \approx 4$.
Given that nearby shakers tend to be aligned: head-to-head for pullers, side-by-side for pushers, as can be deduced from the angular distribution in the inset of \fig{fig:g(r)}(a) \citep{Ge_Elfring_2025}, and that pullers tend to orient in the flow direction whereas pushers perpendicular to it at intermediate Pe (cf.~\fig{fig:map-ori}), we infer that the combined effect is for two nearly touching shakers (pullers or pushers) to form a \emph{doublet} with their activity-induced flow opposing that of the shear \emph{together}.
This is another effect of hydrodynamic interactions in enhancing the relative viscosity.

We have checked that the tendency to form doublets tilted in the flow direction also exists in suspensions of ABPs, but the tendency is generally weaker and it only occurs if the intrinsic persistence length $\ell_{p,0} \gtrapprox 1$; see \fig{fig:map-ori-r}.
A subtle effect of this alignment is that a pair of colliding ABPs generates a pusher-type force dipole through their excluded-volume interactions (\ie~the $\bm F_\text{ext}$ in our simulations) \citep{Edmond_mips}, which affects distant particles and \emph{assists} the shear flow.
Indeed, ABPs tend to collide into their neighbors, as can be seen in the biased angular distributions in the inset of \fig{fig:g(r)}(b).
However, both the shear-induced alignment and shear-thinning rheology are rather weak in ABP suspensions, thus it is difficult to determine to what extent these two phenomena are related.

\section{Conclusion}
\label{sec:conclusion}

We have presented a detailed numerical study on the dissipation and microstructure in sheared active suspensions of squirmers.
Focusing on individually immotile but hydrodynamically interacting squirmers, or shakers, we show how shear enhances the total dissipation but reduces the relative viscosity, corresponding to a shear-thinning rheology, for both puller- and pusher-type shakers.
This rheological response is accompanied by unusual microstructural signatures, characterized by an enhanced nematic order and anisotropic pair correlations, which differ for pullers and pushers.
We further simulate suspensions of self-propelled, neutral squirmers (similar to ABPs), which exhibit weaker shear-thinning, thus highlighting the different effects of activity and motility on the energy expenditure.\footnote{Although we did not study squirmers with both activity and motility, we expect their rheological response to be similar to shakers, since the latter already contain the essential features (persistent motion and hydrodynamically-induced reorientation) of self-propelling squirmers.}
Together with our previous study on the dynamics of the same system \citep{Ge_Brady_Elfring2025}, these results demonstrate the nonlinear interplay of activity and shear, as well as the important effect of hydrodynamic interactions on the microstructure, in suspensions of squirmers.

In our view, active suspensions offer an ideal platform for studying the interaction of various forcing mechanisms in complex fluids, with much of their dynamical and mechanical properties to be explored.
One outstanding question concerns the nature of the superfluid transition, which appears to be related to the emergence of collective motion \citep{Martinez2020}.
Although our simulations of spherical squirmers under simple shear do not display such phenomena, experiments on dilute bacterial suspensions and kinetic theories suggest that elongating the particles may change the response qualitatively.
Whether this is the case in non-dilute suspensions, and if so, how it is affected by hydrodynamic and steric interactions, remain to be studied.

\backsection[Acknowledgements]{We thank John F.~Brady, Shuo Guo, Sankalp Nambiar, Salima Rafa\"i, and Ignacio Pagonabarraga for useful comments or discussions.}

\backsection[Funding]{Z.~Ge acknowledges support from the Swedish Research Council under Grant No.~2021-06669VR, as well as Nanjing University for a starting grant.}

\backsection[Declaration of interests]{The authors report no conflict of interest.}

\backsection[Data availability statement]{The source code that we use to generate the data of this study is openly available in GitHub at https://github.com/GeZhouyang/FSD.}

\backsection[Author ORCIDs]{Z.~Ge, https://orcid.org/0000-0002-4222-012X; G.~J.~Elfring, https://orcid.org/0000-0002-9602-7261.}

\appendix

\section{\xx{Derivation of $\Phi_\text{int}$ in active suspensions of shakers}}
\label{sec:edi_int}

\xx{In our model, a shaker is an active sphere with a prescribed slip, $\bm u_\text{s} = -a\mathsfbi E_\text{s}\bm\cdot\bm n$, where $a$ is its radius, $\mathsfbi E_\text{s}$ is given by Eq.~\eqref{eq:Es}, and $\bm n$ is the unit normal vector pointing into the fluid.
With $\bm\sigma\bm\cdot\bm n$ the local traction exerted by the fluid on the particle, and the particle free of torque and subject to the interparticle force, $\bm F_\text{ext} = -\bm F_\text{H}$, the rate at which the particle does work on the fluid is
\begin{equation}
  -\oint (\bm\sigma\bm\cdot\bm n)\bm\cdot(\bm U + a\bm\Omega\times\bm n + \bm u_\text{s}) \,\mathrm{d}S
  = \bm F_\text{ext}\bm\cdot\bm U + P ,
  \quad
  P = -\oint (\bm\sigma\bm\cdot\bm n)\bm\cdot\bm u_\text{s} \,\mathrm{d}S ,
\end{equation}
where $\bm U$ and $\bm\Omega$ describe the rigid-body motion of the particle, and $P$ is the input power of the active surface.
Summed over all particles and divided by the suspension volume $V$, the $\bm F_\text{ext}\bm\cdot\bm U$ term corresponds to the rate at which the interaction potential transfer energy, which must vanish on average at steady state (verified in \fig{fig:phi-edi0}(b) inset).
Writing $\bm U = \mathsfbi E_\infty\bm\cdot\bm x + \bm U'$, its affine part equals $-\mathsfb\Sigma_\text{P}\bm:\mathsfbi E_\infty$, so that on average $\frac{1}{V}\sum\bm F_\text{ext}\bm\cdot\bm U' = \mathsfb\Sigma_\text{P}\bm:\mathsfbi E_\infty$, which is $\Phi_\text{ext}$ minus the pure fluid dissipation, \Eq{eq:edi_ext}.
The total dissipation is therefore $\Phi = \Phi_\text{ext} + \Phi_\text{int}$, where $\Phi_\text{int} \equiv n\langle P\rangle$.
}

\xx{Since $\mathsfbi E_\text{s}$ is symmetric and traceless, we can rewrite $P$ compactly as
\begin{equation}\label{eq:P_Str}
  P = \mathsfbi S_\text{tr}\bm:\mathsfbi E_\text{s} ,
\end{equation}
where
\begin{equation}\label{eq:Str}
  \mathsfbi S_\text{tr} = \oint \Big[ \tfrac12 \big( a\bn (\bm\sigma\bm\cdot\bm n) + (\bm\sigma\bm\cdot\bm n)a\bn \big) 
  - \tfrac13 \big( a\bn \bm\cdot\bm\sigma\bm\cdot\bm n \big)\mathsfbi I \Big] \mathrm{d}S
\end{equation}
is the symmetric and traceless first moment of the traction about the particle center.
In general, $\mathsfbi S_\text{tr} \ne \mathsfbi S_\text{H}$, because $\mathsfbi S_\text{H}$ contains an additional velocity moment \citep{elfring2022active},
\begin{equation}\label{eq:SH_Str}
  \mathsfbi S_\text{H} = \mathsfbi S_\text{tr} - \eta\oint (\bm u\bm n + \bm n\bm u) \,\mathrm{d}S .
\end{equation}
The rigid-body part of $\bm u$ does not contribute to the velocity moment by symmetry.
Only the slip remains, and with $\oint\bm n\bm n\,\mathrm{d}S = \frac{4}{3}\pi a^2\mathsfbi I$ on the particle surface, we have
\begin{equation}
  -\eta\oint (\bm u_\text{s}\bm n + \bm n\bm u_\text{s}) \,\mathrm{d}S
  = \eta a \big( \mathsfbi E_\text{s}\bm\cdot\oint\bm n\bm n\,\mathrm{d}S + \oint\bm n\bm n\,\mathrm{d}S\bm\cdot\mathsfbi E_\text{s} \big)
  = \frac{8}{3}\pi\eta a^3\mathsfbi E_\text{s} .
\end{equation}
Hence,
\begin{equation}\label{eq:SH_Str1}
  \mathsfbi S_\text{H} = \mathsfbi S_\text{tr} + \frac{8}{3}\pi\eta a^3\mathsfbi E_\text{s} .
\end{equation}
Note that, the difference between $\mathsfbi S_\text{H}$ and $\mathsfbi S_\text{tr}$ depends only on the particle's slip, not the configuration of the suspension or the externally imposed flow.
}

\xx{Averaging Eq.~\eqref{eq:P_Str} over all particles and using Eq.~\eqref{eq:SH_Str1} with $n\frac{4}{3}\pi a^3 = \phi$, we thus obtain
\begin{equation}\label{eq:edi_int_app}
  \Phi_\text{int} = n\langle\mathsfbi S_\text{tr}\bm:\mathsfbi E_\text{s}\rangle
  = n\langle\mathsfbi S_\text{H}\bm:\mathsfbi E_\text{s}\rangle - 2\eta\phi\,\langle\mathsfbi E_\text{s}\bm:\mathsfbi E_\text{s}\rangle ,
\end{equation}
which is Eq.~\eqref{eq:edi_int}.
}

\section{Size ($N$) dependence of $\eta_r$ at small Pe}
\label{sec:N}

\begin{figure*}
  \centering
  \includegraphics[width=\columnwidth]{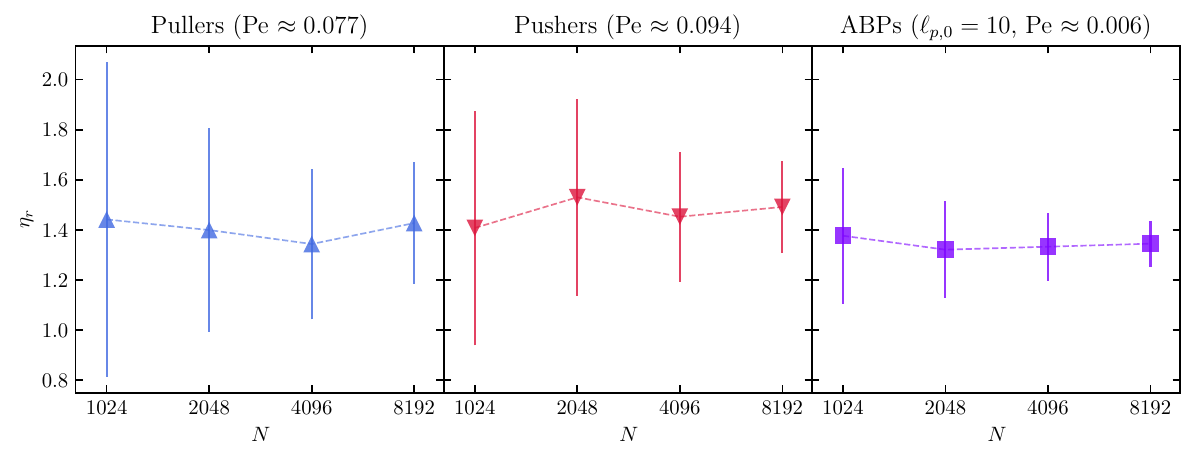}
  \begin{picture}(0,0)
  \end{picture}
  \caption{Size dependence of the relative viscosity for pulling shakers, pushing shakers, and ABPs at very small shear rates and $\phi=10\%$.}
  \label{fig:eta-N}
\end{figure*}

In \S \ref{sec:rheo} of the main text, we mentioned that the relative viscosities for shakers at Pe $\ll 1$ and ABPs at Pe $\ll 0.1$ have large fluctuations.
\Fig{fig:eta-N} illustrates this behavior for a few representative cases at $\phi=10\%$.
As the number of particles $N$ increases from 1024 to 8192, the fluctuations reduce significantly, whereas the averaged relative viscosity $\eta_r$ remains roughly the same.
Furthermore, we note that the $\eta_r$ for pullers, pushers, and ABPs, respectively, at $N=8192$ have similar values to their first data points in \fig{fig:visc-phi10}(a), where the Pe are about 10 times higher.
Specifically, $\eta_r= 1.396 \pm 0.078$ for pulling shakers at Pe $\approx 0.775$, $\eta_r= 1.451 \pm 0.062$ for pushing shakers at Pe $\approx 0.936$, and $\eta_r= 1.336 \pm 0.028$ for ABPs ($\ell_{p,0}/a=10$) at Pe $\approx 0.060$.
Our results thus suggest that the relative viscosities presented in \fig{fig:visc-phi10}(a) have converged to their limiting values as Pe $\to 0$.

\bibliographystyle{jfm}
\bibliography{main}

\end{document}